%% file: main.tex
\documentclass[letter,journal,twoside]{IEEEtran}

\usepackage[pdftex]{graphicx}
\usepackage{amsmath}
\usepackage{algorithmic}
\usepackage{siunitx}
\usepackage{mhchem}
\usepackage{authblk}
\usepackage{booktabs}
\usepackage{multirow}
\usepackage{fancyhdr}
\usepackage{cite}
\usepackage[many]{tcolorbox}
\usepackage{lipsum}
\usepackage{RobStd}
\usepackage{tikz}
\usepackage{rotating}
\usepackage{placeins}
\usepackage{soul}

\soulregister\cite7
\soulregister\ref7
\soulregister\pageref7
\soulregister\xspace7

\renewcommand{\hl}[1]{#1}

\ifCLASSOPTIONcompsoc
    \usepackage[caption=false,font=normalsize,labelfont=sf,textfont=sf]{subfig}
\else
    \usepackage[caption=false,font=footnotesize]{subfig}
\fi

\input{comments}

\setboolean{showcomments}{false}

\begin{document}
\renewcommand{\arraystretch}{1.1}

\title{A 185\,TOPS/W/mm$\mathrm{^\mathbf{2}}$ Bayesian Inference Engine with 640\,aJ, Write-Free FeFET GRNG for Uncertainty-Aware Aerial Search and Rescue}

\author{Zephan M. Enciso,~\IEEEmembership{Student Member,~IEEE,}
Xuezhong Niu,~\IEEEmembership{Student Member,~IEEE,}
Xingtian Wang,~\IEEEmembership{Student Member,~IEEE,}
Mohammad Mehdi Sharifi,~\IEEEmembership{Member,~IEEE,}
Subhasish Mukherjee,~\IEEEmembership{Student Member,~IEEE,}
Likai Pei,~\IEEEmembership{Student Member,~IEEE,}
Halid Mulaosmanovic,~\IEEEmembership{Member,~IEEE,}
Stefan Duenkel,~\IEEEmembership{Member,~IEEE,}
Sven Beyer,~\IEEEmembership{Member,~IEEE,}
Michael Niemier,~\IEEEmembership{Member,~IEEE,}
Kai Ni,~\IEEEmembership{Member,~IEEE,}
Ningyuan Cao,~\IEEEmembership{Member,~IEEE}
\thanks{Manuscript received Mmm DD, YYYY; revised Mmm DD YYYY.}%
\thanks{Zephan Enciso, Xuezhong Niu, Xingtian Wang, Mohammad Mehdi Sharifi, Subhasish Mukherjee, Likai Pei, Michael Niemier, Kai Ni, and Ningyuan Cao are with the University of Notre Dame du Lac, Notre Dame, IN, USA.}%
\thanks{Halid Mulaosmanovic, Stefan Duenkel, and Sven Beyer are with GlobalFoundries Fab1, 01109 Dresden, Germany.}
\thanks{Color versions of one or more figures in this article are available at \href{https://doi.org/00.0000/JSSC.0000.0000000}{https://doi.org/00.0000/XXXXX.0000.0000000}}%
\thanks{Digital Object Identifier 00.0000/XXXXX.0000.0000000}%
}

\markboth{IEEE Transactions on Circuits and Systems for Artificial Intelligence,~Vol.~XX, No.~X, Mmm~YYYY} {Enciso \MakeLowercase{\textit{et al.}}: A Bayesian Inference Engine for Uncertainty-Aware Aerial Search and Rescue}

\IEEEpubid{0000--0000/00\$00.00~\copyright~2026 IEEE}

\maketitle

\definecolor{green5}{HTML}{26a269}
\definecolor{blue5}{HTML}{1a5fb4}
\definecolor{orange3}{HTML}{ff7800}
\definecolor{red3}{HTML}{e01b24}
\definecolor{red4}{HTML}{c01c28}
\definecolor{purple4}{HTML}{813d9c}

\begin{abstract}

The success of an aerial search and rescue (SAR) mission depends on locating the victim within a rapidly shrinking survival window.
Deterministic deep learning models often produce overconfident predictions in uncertain environments, leading to verification maneuvers that decrease search coverage and increase rescue delay.
Bayesian neural networks (BNNs) provide uncertainty-aware decision-making that can flag potential false positives,
but their sampling overhead results in high power consumption that reduces flight endurance.
Emerging memory devices can address this via stochastic programming, but such approaches typically suffer from poor scalability (requiring large devices) or limited endurance (requiring frequent write operations).

This work presents a technology-scalable solution: a write-free, central limit theorem-based Gaussian random number generator (CLT-GRNG) embedded within a ferroelectric FET (FeFET) compute-in-memory macro.
By summing currents from a randomly selected subset of minimum-sized, programmed-once FeFETs, the proposed architecture eliminates energy- and wear-intensive write cycles during inference and the need for large, precision-tuned devices.
The CLT-GRNG consumes just 640\,aJ per sample---a 560$\mathbf{\times}$ gain over prior BNN accelerators---and the CIM tile attains 185\,TOPS/W/mm$^\mathrm{\textbf{2}}$ \hl{area-normalized} compute efficiency.
This approach enables robust, uncertainty-aware SAR detection without compromising flight endurance, which shortens rescue delay and ultimately increases victim survivability.

\end{abstract}


\input{00_Introduction}
\input{01_Background}
\input{02_GRNG}
\input{03_Tile}
\input{04_Evaluation}
\input{05_Conclusion}

\section*{Acknowledgments}

This work is partially funded by the European Union within ``NextGeneration EU,'' by the Federal Ministry for Economic Affairs and Energy (BMWE) on the basis of a decision by the German Bundestag and by the State of Saxony with tax revenues based on the budget approved by the members of the Saxon State Parliament in the framework of ``Important Project of Common European Interest---Microelectronics and Communication Technologies'', under the project name ``EUROFOUNDRY.''
This work is also supported in part by SUPREME, one of seven centers in JUMP 2.0, a Semiconductor Research Corporation (SRC) program sponsored by DARPA.

\bibliographystyle{IEEEtran}
\bibliography{references}

\begin{IEEEbiography}[{\includegraphics[width=1in,height=1.25in,clip,keepaspectratio]{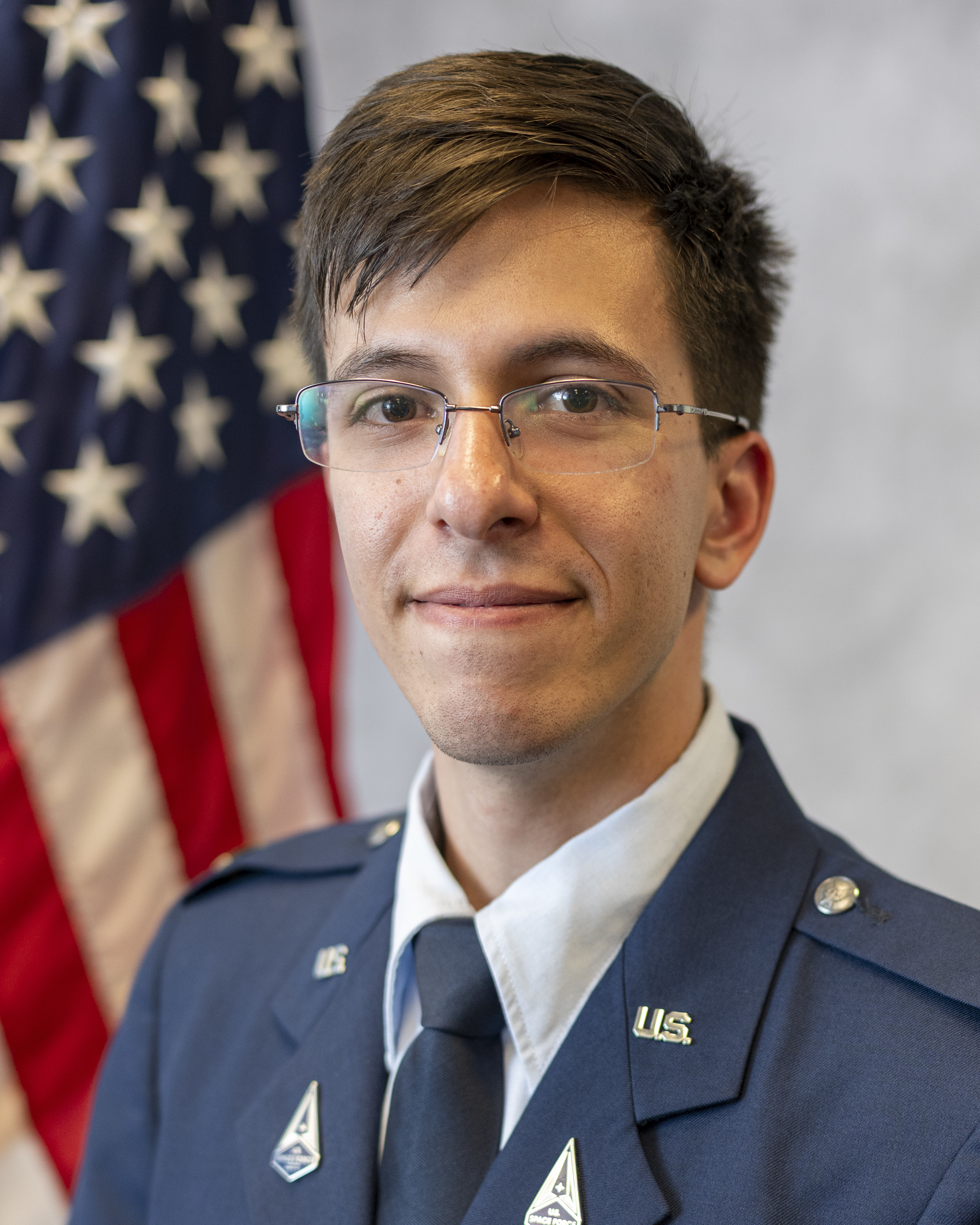}}]{Zephan M. Enciso}
received his B.S.\ in Computer Engineering and B.S.\ in Electrical Engineering from the University of Notre Dame, Notre Dame, Indiana before returning to the University of Notre Dame to pursue his Ph.D.
His research interests include the deployment of edge inference in safety-critical, resource-constrained systems, hardware acceleration of uncertainty-aware artificial intelligence, and novel devices, circuits, and architectures for machine learning.
Z.\ M.\ Enciso was a Department of Defense National Defense Science and Engineering Graduate Fellowship recipient, a Design Automation Conference Young Fellow, and a recipient of the Jack and Mary Ann Remick Fellowship in Engineering.
He was also inducted into Sigma Xi, The Scientific Research Honors Society, the IEEE Eta Kappa Nu Honors Society, and the ACM Upsilon Pi Epsilon Honors Society.

\end{IEEEbiography}

\begin{IEEEbiography}[{\includegraphics[width=1in,height=1.25in,clip,keepaspectratio]{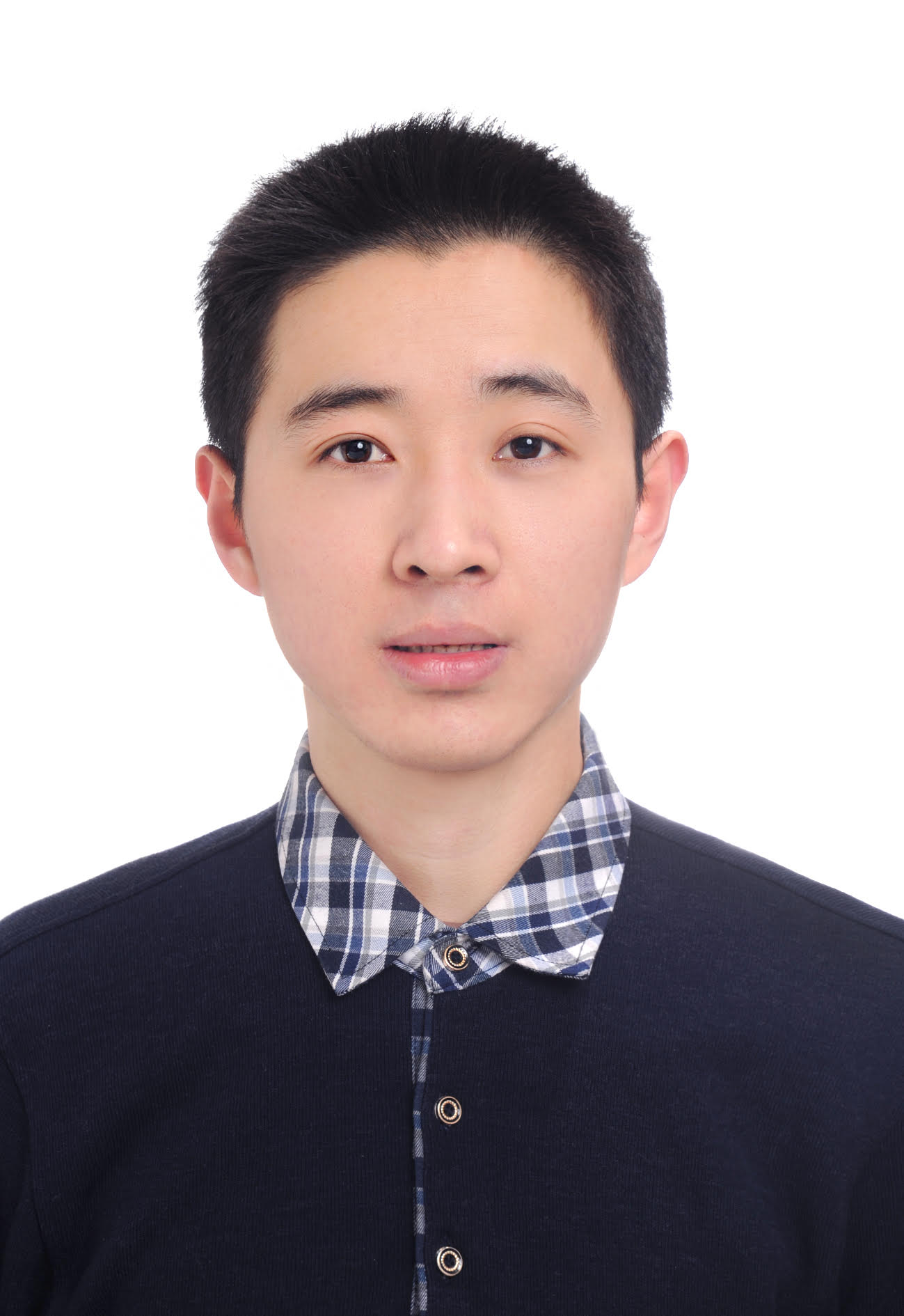}}]{Xuezhong Niu}
is currently pursuing his Ph.D.\ degree in the Department of Electrical Engineering, University of Notre Dame, Notre Dame, IN, USA.
He received his M.S.\ degree in electrical engineering from East China Normal University, China, in 2023.
In 2024, he joined Notre Dame as a Ph.D.\ student.
His research interest focuses on emerging ferroelectric field-effect transistors for monolithic 3-D integration.
\end{IEEEbiography}

\begin{IEEEbiography}[{\includegraphics[width=1in,height=1.25in,clip,keepaspectratio]{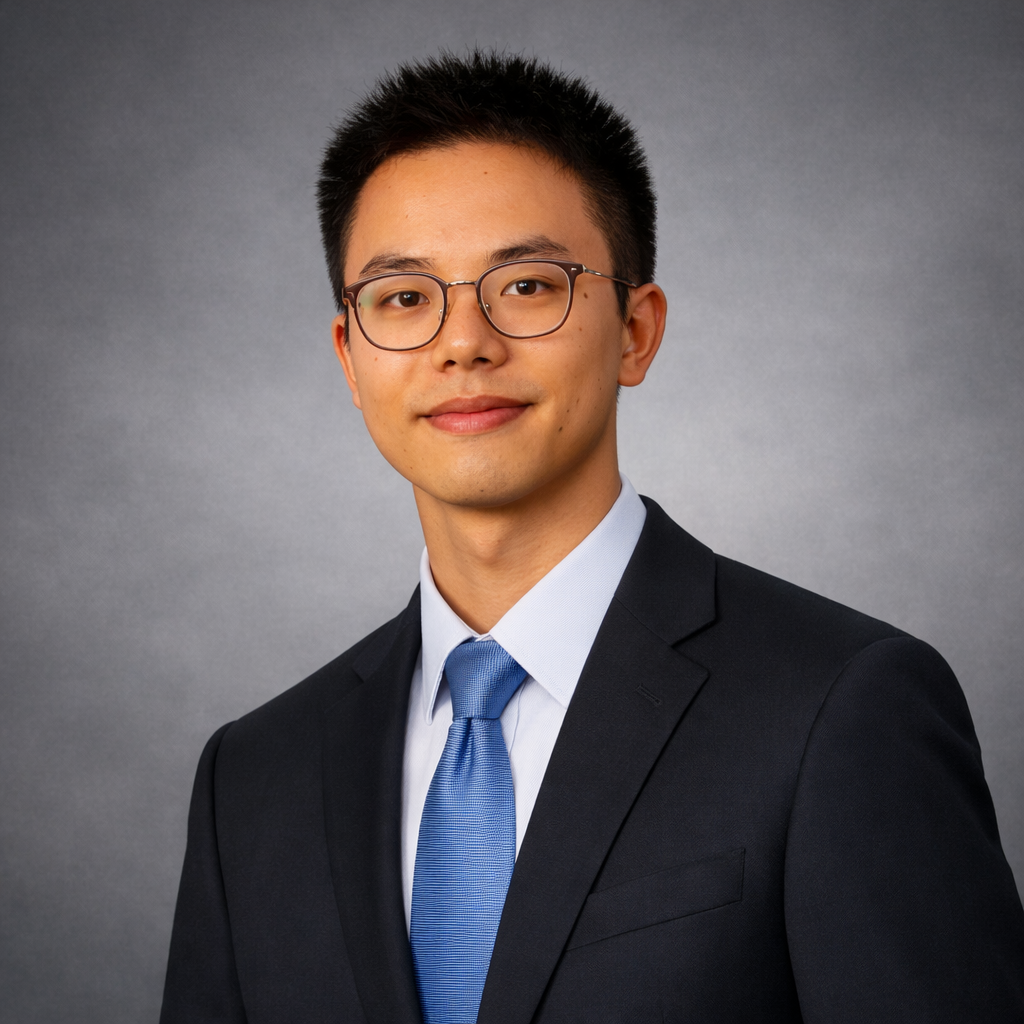}}]{Xingtian Wang}
received the bachelor’s degree in electronic information engineering from the University of Electronic Science and Technology of China, Chengdu, China, in 2024.
He is currently pursuing the Ph.D.\ degree in electrical engineering at the University of Notre Dame, Notre Dame, IN, USA.
His research focuses on oxide semiconductor devices, advanced CMOS devices, with applications in emerging electronic systems.

\end{IEEEbiography}

\begin{IEEEbiography}[{\includegraphics[width=1in,height=1.25in,clip,keepaspectratio]{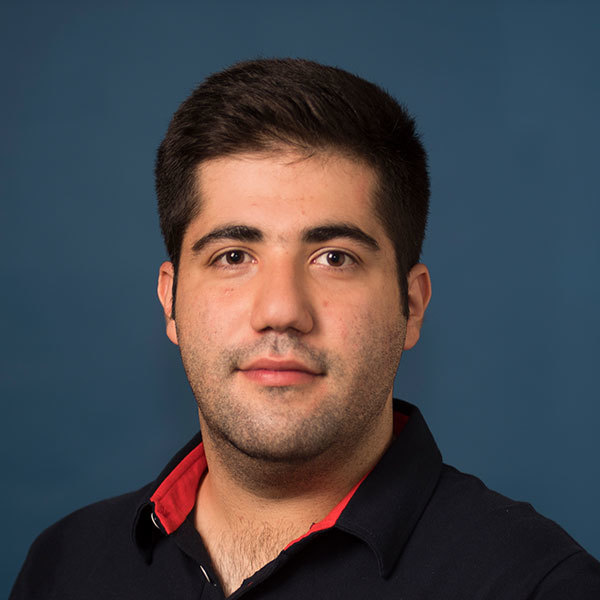}}]{Mohammad Mehdi Sharifi}
earned his bachelor's degree in computer engineering from Shahid Beheshti University, Tehran, Iran, in 2017.
He recently obtained his Ph.D.\ degree at Notre Dame, under the joint supervision of Dr.\ X.\ Sharon Hu and Dr.\ Michael Niemier.
He is presently working as a post-doctoral researcher at Notre Dame.
His research focuses on low-power circuit design, applications for beyond-CMOS technologies, and in-memory computing.
Specifically, he is interested in designing and benchmarking circuits and architectures that leverage the unique properties of beyond-CMOS technologies, such as FeFETs.
In recognition of his contributions, he was nominated for the Best Paper Award at DATE 2021.
\end{IEEEbiography}

\begin{IEEEbiography}[{\includegraphics[width=1in,height=1.25in,clip,keepaspectratio]{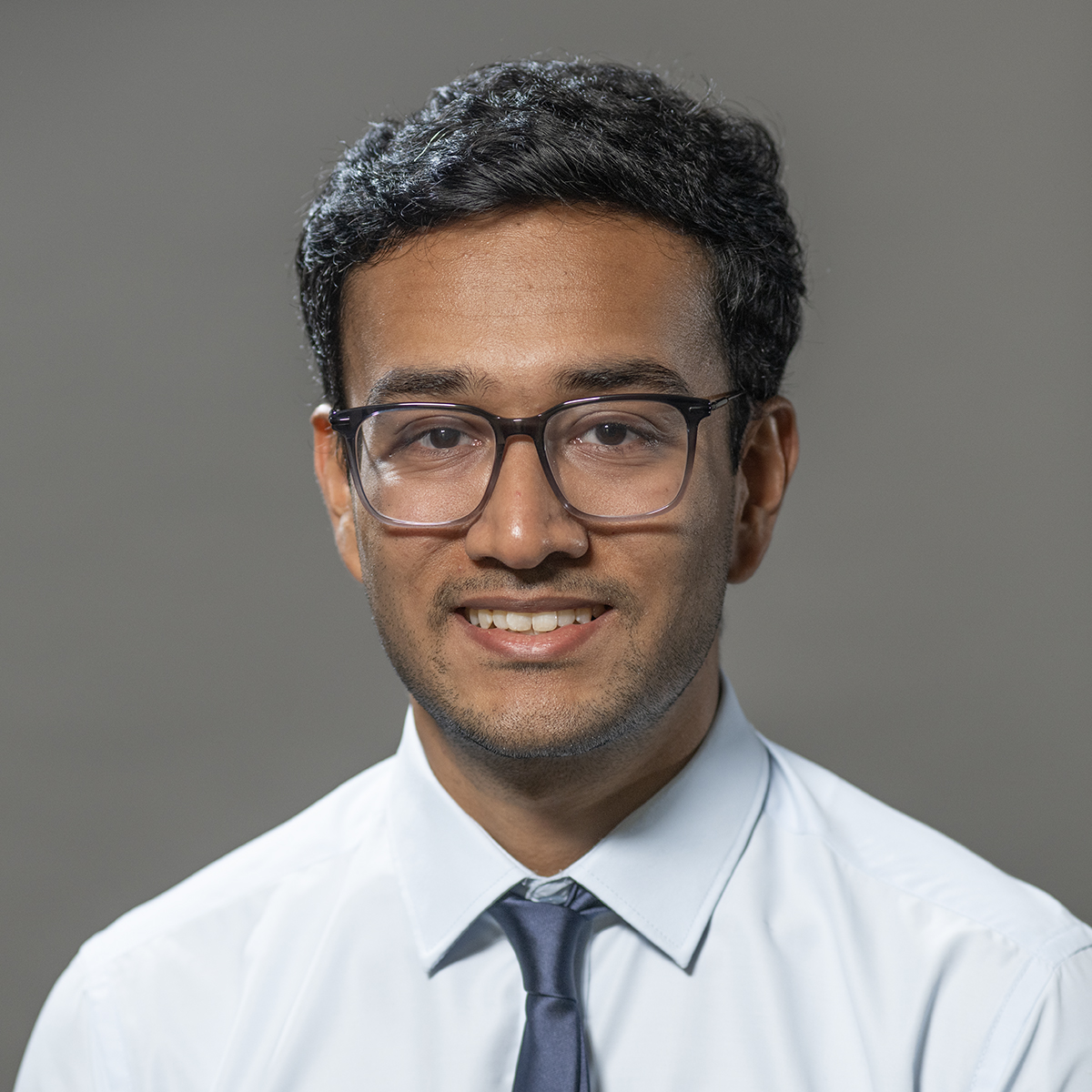}}]{Subhasish Mukherjee}
received his B.Tech.\ degree in Electrical Engineering from JGEC, West Bengal, India, and his M.Tech.\ degree in Electrical Engineering from the Indian Institute of Science, Bangalore, India, in 2020.
He subsequently worked as a SoC Design Engineer at Intel Corporation, where he contributed to the design and optimization of advanced processor architectures.
In August 2023, he joined the University of Notre Dame as a Ph.D.\ student in Computer Science.
His research focuses on accelerating large-scale AI applications, including transformers and recommender systems, using emerging in-memory computing technologies and hardware–algorithm co-design.
\end{IEEEbiography}

\begin{IEEEbiography}[{\includegraphics[width=1in,height=1.25in,clip,keepaspectratio]{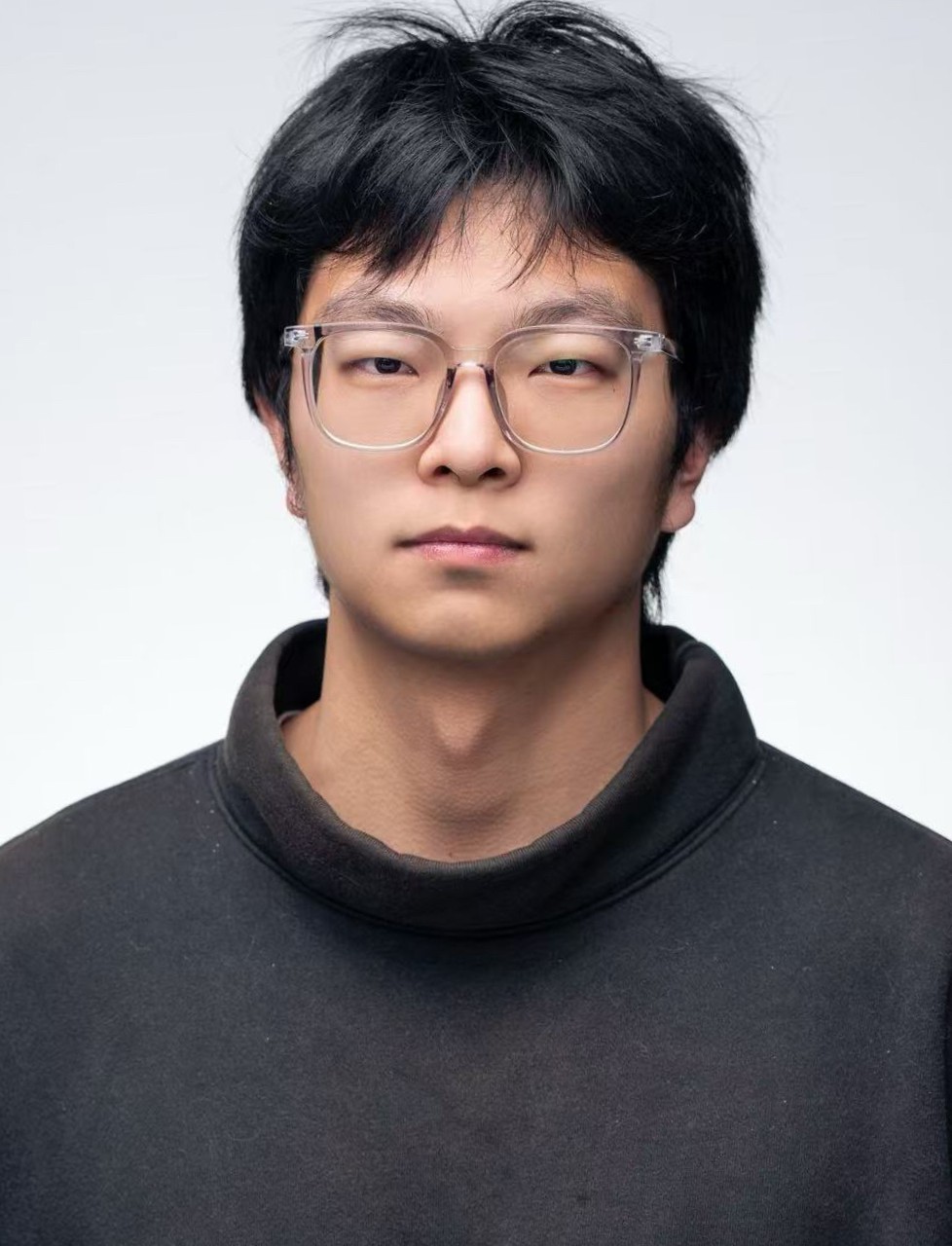}}]{Likai Pei}
received his B.S.\ degree in Electrical Engineering from Nanjing University of Information Science and Technology, Nanjing, China, in 2023.
He completed his M.S.\ studies and began pursuing the Ph.D.\ degree in electrical engineering at the University of Notre Dame, Notre Dame, IN, USA, in 2024.
His research interests include analog/mixed-signal circuit design, explainable artificial intelligence, and low-power neural networks integrated with emerging devices for edge computing.

\end{IEEEbiography}

\begin{IEEEbiography}[{\includegraphics[width=1in,height=1.25in,clip,keepaspectratio]{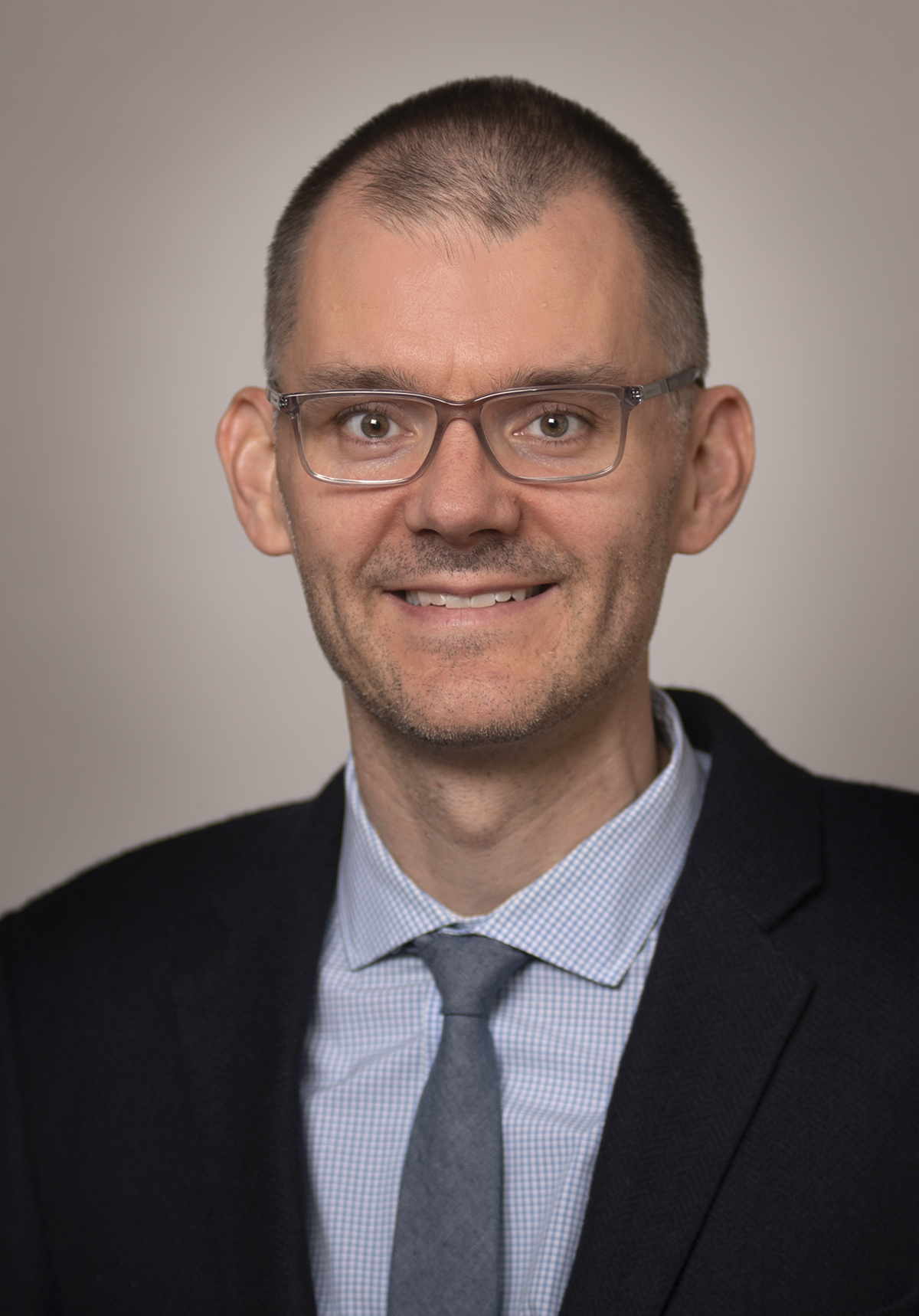}}]{Michael Niemier}
is currently a Professor at the University of Notre Dame.
His research interests include designing, facilitating, benchmarking, and evaluating circuits and architectures based on emerging technologies.
Currently, Niemier's research efforts are based on new transistor technologies, as well as devices based on alternative state variables such as spin.
He is the recipient of an IBM Faculty Award, the Rev.\ Edmund P.\ Joyce, C.S.C.\ Award for Excellence in Undergraduate Teaching at the University of Notre Dame, and best paper awards such as at ISLPED.
Niemier has served on numerous technical program committees for design related conferences (including DAC, DATE, ICCAD, etc.), and has chaired the emerging technologies track at DATE, DAC, and ICCAD.
He is an associate editor for IEEE Transactions on Nanotechnology, as well as the ACM Journal of Emerging Technologies in Computing.
Niemier is also an avid runner and enjoys traveling with his family.

\end{IEEEbiography}

\begin{IEEEbiography}[{\includegraphics[width=1in,height=1.25in,clip,keepaspectratio]{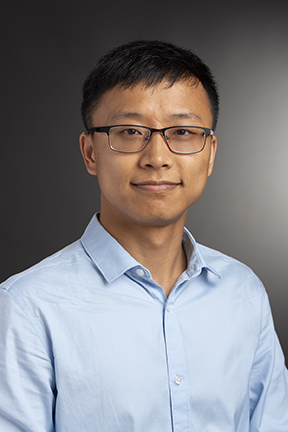}}]{Kai Ni}
received the B.S.\ degree in Electrical Engineering from University of Science and Technology of China, Hefei, China in 2011, and Ph.D.\ degree of Electrical Engineering from Vanderbilt University, Nashville, TN, USA in 2016 by working on advanced electronics for space applications.
Since then, he became a postdoctoral associate at University of Notre Dame, working on ferroelectric devices for nonvolatile memory and novel computing paradigms.
He joined Electrical and Microelectronic Engineering department at Rochester Institute of Technology as an assistant professor in 2019.
After four years, he joined University of Notre Dame as an assistant professor in the Electrical Engineering department in 2023.
His current research interests lie in nanoelectronic devices empowering next generation storage and computing hardware.
He served as technical program committee of several conferences, including IEDM, IRPS, DRC, DAC, DATE.
\end{IEEEbiography}

\begin{IEEEbiography}[{\includegraphics[width=1in,height=1.25in,clip,keepaspectratio]{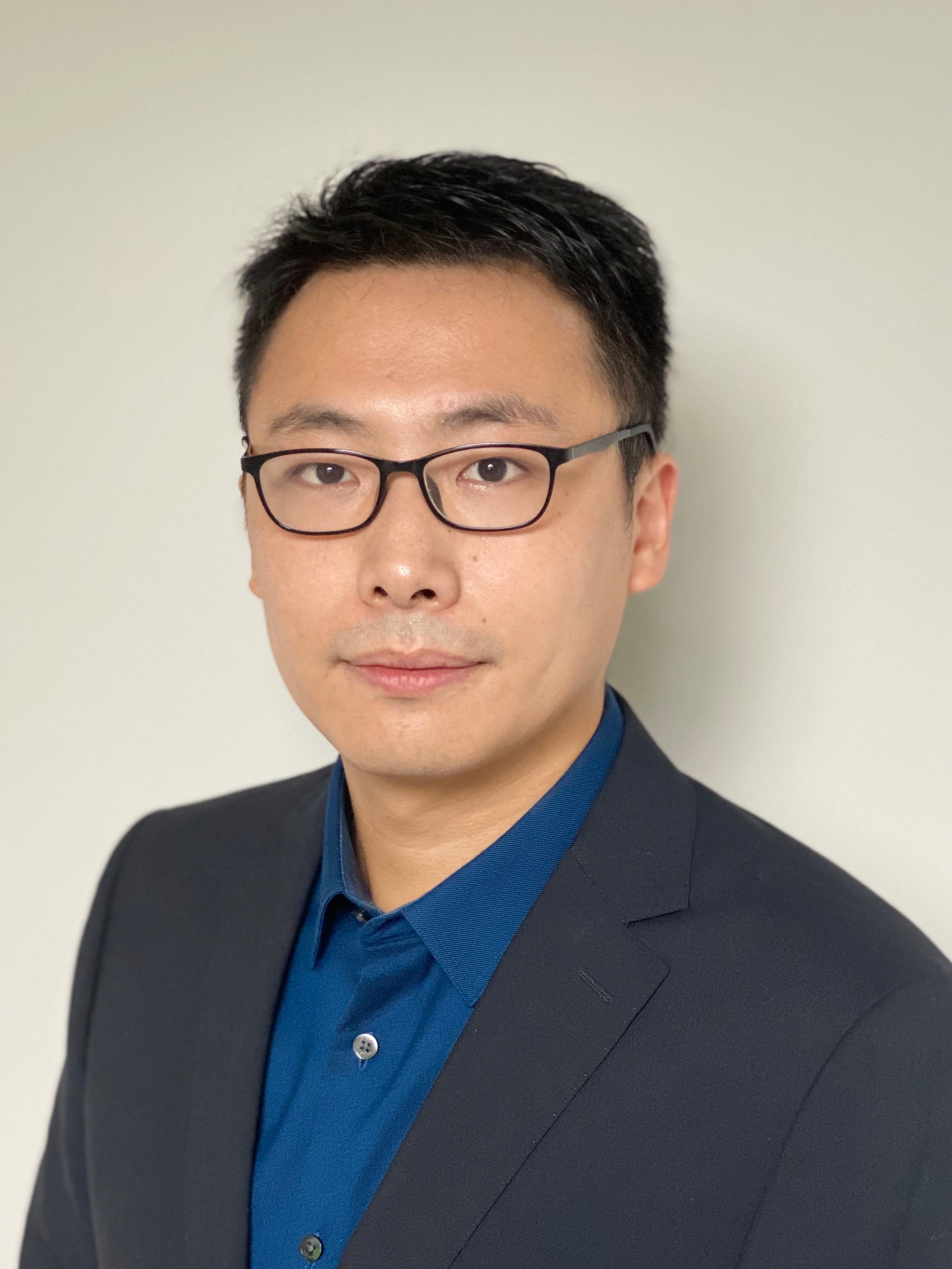}}]{Ningyuan Cao}
received the bachelor’s degree from Shanghai Jiao Tong University, Shanghai, China, in 2013, the master’s degree from Columbia University, New York City, NY, USA, in 2015, and the Ph.D.\ degree in integrated circuit and algorithm design for edge intelligence from the Georgia Institute of Technology, Atlanta, GA, USA, in 2020.
He is currently an Assistant Professor with the Department of Electrical Engineering at the University of Notre Dame, Notre Dame, IN, USA.
Prior to joining Notre Dame, he was a Research Associate with the IBM Thomas J. Watson Research Center, Yorktown Heights, NY, USA.
His research work has been published and presented in top-tier conferences including ISSCC, VLSI Symposium, DAC, IMS, CICC, and ICCAD, as well as leading journals such as the IEEE Journal of Solid-State Circuits, IEEE Internet of Things Journal, IEEE Transactions on Industrial Electronics, and IEEE Transactions on Circuits and Systems—I: Regular Papers.
He is also an inventor on multiple U.S.\ patents and patent applications.
His research interests span analog and mixed-signal circuits, energy-efficient digital architectures, and system-on-chip design for edge intelligence, with a particular emphasis on trustworthy AI hardware, including uncertainty-aware, privacy-preserving, and robust mixed-signal computing systems for resource-constrained and safety-critical applications.
\end{IEEEbiography}

\end{document}

%% file: comments.tex
\usepackage{ifthen}
\usepackage{amssymb}
\usepackage{xcolor}

\newboolean{showcomments}
\setboolean{showcomments}{true}

\makeatletter
\newcommand{\mynote}[3]{%
  \ifthenelse{\boolean{showcomments}}{%
   \fbox{\bfseries\sffamily\scriptsize#1}%
   {\small$\blacktriangleright$\textsf{\emph{\color{#3}{#2}}}$\blacktriangleleft$}}%
  {%
   \@bsphack
   \@esphack
  }%
}
\makeatother

%% file: 00_Introduction.tex
\section{Introduction}
\label{sec:intro}

\begin{figure}[t]
    \centering
	\includegraphics[width=0.9\columnwidth]{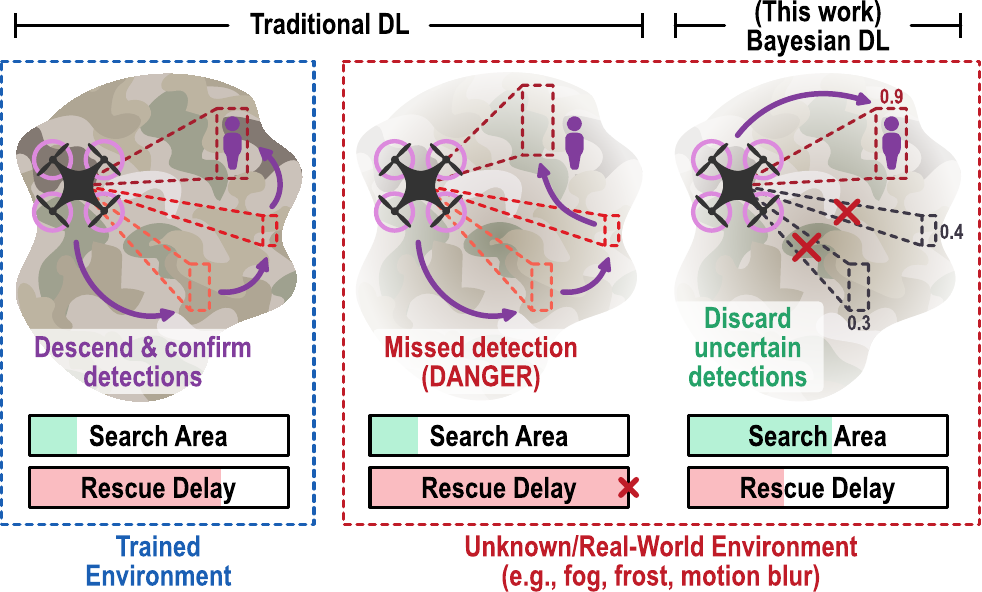}

	\caption{In a fully autonomous SAR fleet, a UAS must deviate from its search path and descend to verify each detection; these confirmations are costly in terms of battery and time. Furthermore, real-world SAR environments are unpredictable and often differ from training data, leading to missed detections. Bayesian DL addresses these issues via predictive uncertainty, enabling the system to discard low-confidence predictions---thereby maintaining altitude, maximizing search coverage, and locating victims faster. They are also more resilient to out-of-distribution inputs, retaining more detection accuracy in challenging environments.}
    \label{fig:motivation}

\end{figure}

\IEEEPARstart{A}{erial} search and rescue (SAR) operations are critical, time-sensitive missions where unmanned aircraft systems (UASs) are deployed to locate missing persons in wilderness or disaster environments.
The efficacy of these operations is governed by a ``golden window'' of survival, which degrades rapidly due to exposure, injury, or hazardous weather conditions.
Manual review of aerial footage is labor-intensive and prone to operator fatigue, while human piloting limits the number of aircraft that can be fielded simultaneously~\cite{abdelnabi2024human, lyu2023unmanned}.
Consequently, there has been a paradigm shift toward autonomous, multi-UAS clusters capable of scanning vast areas rapidly; these systems can improve search times by over \qty{160}{\percent} but require robust, real-time onboard detection~\cite{ewers2025deep, sambolek2021automatic}.
Furthermore, processing data at the edge is essential, as communication bandwidth in disaster zones is often unreliable or nonexistent, precluding cloud-based analysis~\cite{zaidalkilani2025ai}.

\IEEEpubidadjcol Deep learning (DL) models have become the standard for automated victim detection.
However, standard DL models are deterministic and frequently suffer from overconfidence~\cite{ras2022explainable}, classifying image artifacts as victims with high probability.
In a fully autonomous deployment, a drone must deviate from its flight path and descend to verify each high-confidence detection.
False positives therefore initiate costly confirmation loops that waste battery power and flight endurance, ultimately shrinking the searchable area and increasing rescue delay---critical factors that directly reduce the likelihood of finding victims before the survival window closes.
Moreover, standard models lack robustness to out-of-distribution (OOD) inputs.
If a UAS encounters environmental conditions differing from its training set, performance degrades catastrophically because the model forces unknown inputs into known classes, potentially missing victims entirely~\cite{hullermeier2021aleatoric}.

Bayesian neural networks (BNNs) offer a compelling solution to these reliability issues by replacing fixed weights with probability distributions.
This enables the system to quantify predictive uncertainty.
By filtering out detections with high uncertainty, such as sensor noise or environmental anomalies, a BNN-equipped UAS can disregard low-confidence predictions.
\hl{Fig.~\ref{fig:motivation} shows how this uncertainty-informed selective processing preserves energy for verifying valid targets,} thereby accelerating victim localization to maximize the probability of a successful live rescue.
Despite these advantages, BNNs are computationally expensive; traditional hardware struggles with the generation and movement of massive quantities of random samples within the tight power constraints of UASs.

One promising avenue to more efficiently implement BNNs is to exploit the intrinsic stochasticity of emerging memory devices (EMDs) to perform multi-bit, in-memory RNG~\cite{sebastian2022two}.
Co-locating GRNG with weight storage can dramatically reduce data movement and improve energy efficiency.
Unfortunately, single-device distribution representations rely on multi-level programming, which does not scale well with technology node shrinkage. 
Fundamental device properties, such as the number of grains in a ferroelectric device~\cite{kazemi2022achieving} or the crystal count of a phase-change memory~\cite{deng2020comprehensive}, constrain the number of reliably distinguishable conductance states, forcing designers to either sacrifice precision, compromise device endurance with write-verify loops, or tolerate the overhead of larger devices~\cite{sharifi2021application}---thus limiting maximum attainable efficiency.

This work resolves these scalability issues by embedding a central limit theorem-based GRNG (CLT-GRNG) directly within the memory array.
By summing currents from several minimum-sized FeFETs, the CLT-GRNG produces a high-quality Gaussian distribution regardless of the individual process variations of highly-scaled devices.
Crucially, the FeFETs within the CLT-GRNG are programmed to a high-entropy state only once; the system generates random samples by reading from a dynamically selected subset of this array.
This write-free approach eliminates the latency and power consumption associated with rewriting devices and dramatically increases device longevity.
Consuming only \qty{640}{\atto\joule} per sample (including selection logic), the CLT-GRNG delivers a $560\times$ improvement in energy efficiency over state-of-the-art (SOTA) GRNG~\cite{liu2025va}.
Integrated into a 185\,TOPS/W/mm$^\mathrm{2}$ FeFET BNN accelerator, this solution enables truly scalable, energy-efficient stochastic computation, providing a reliable foundation for next-generation autonomous SAR systems.

%% file: 01_Background.tex
\section{Background}

\begin{figure}
    \centering
    \includegraphics[width=\columnwidth]{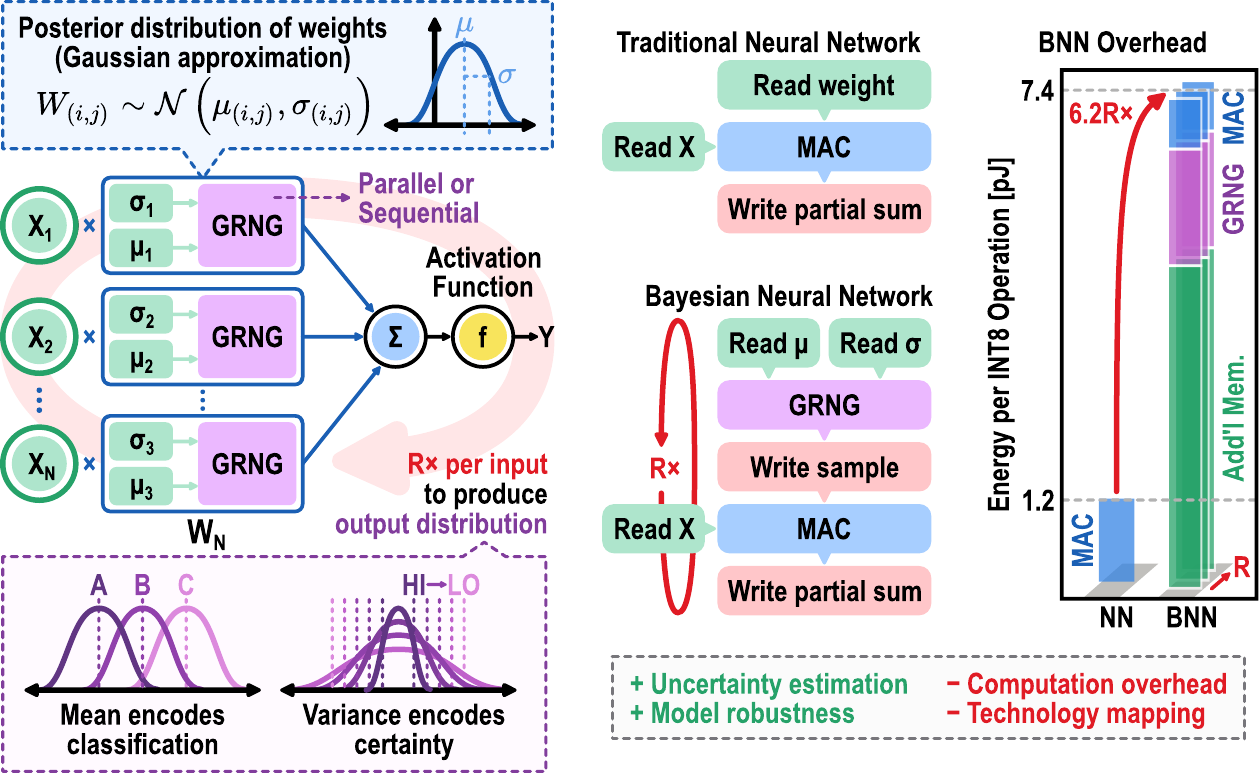}

    \caption{Conventional BNN neuron.  The GRNG generates samples from a $\mathcal{N}\left(\mu, \sigma\right)$ distribution for each weight, requiring storage of both $\mu$ and $\sigma$ and multiple memory access ($R\times$) per inference iteration.  This dramatically increases computational overhead, as shown in the plot on the right.}
	\label{fig:overhead}

\end{figure}

\subsection{Bayesian Neural Networks}

To enable a probabilistic interpretation of outputs, BNNs treat each weight as a random variable rather than a fixed scalar~\cite{goan2020bayesian}.
Hardware deployments for edge systems typically approximate these arbitrary posterior distributions with Gaussian distributions via variational inference (VI)~\cite{hoffman2013stochastic, blei2017variational}, per Eq. (\ref{eq:variational}):

\begin{equation}
    \label{eq:variational}
	P\left(\mathbf{W}\, |\, \mathbf{X}, \mathbf{Y}\right) \approx \mathcal{N} \left( \mathbf{W}\, |\, \mathbf{\mu}, \mathbf{\sigma} \right)
\end{equation}

\noindent Here, $\mathbf{\mu}$ and $\mathbf{\sigma}$ denote the mean and covariance matrix of the Gaussian distribution, respectively.
Replacing each deterministic weight with a Gaussian distribution yields a network that outputs a classification \textit{distribution} after $R$ inference iterations.
The variance of this output distribution encodes model uncertainty arising from noisy inputs (aleatoric uncertainty) or out-of-distribution inputs (epistemic uncertainty)~\cite{hullermeier2021aleatoric}.
However, this reliance on stochastic sampling significantly amplifies hardware resource demands; \hl{in Fig.~\ref{fig:overhead}, the BNN's additional operations result in $6.2\times$ more energy per \texttt{INT8} for each sampling pass}.
Furthermore, the \textit{quality} of uncertainty estimates improves with the number of passes ($R$) per input, creating a fundamental trade-off between computational overhead and explainability.

\hl{Despite their theoretically superior robustness and uncertainty quantification, BNNs have seen limited adoption in commercial edge AI solutions due to the prohibitive computational and memory overhead of stochastic sampling~\cite{jospin2022hands}.
Despite these hurdles, academic research has demonstrated the effectiveness of BNNs in safety-critical domains where uncertainty quantification is vital.
In biomedical applications, BNNs have outperformed traditional models in diagnosing Alzheimer's disease~\cite{ferrante2024enabling} and identifying ventricular arrhythmias~\cite{enciso2026350} by deferring uncertain cases.
Furthermore, BNNs have been proposed for safety verification of nonlinear control systems, such as UASs, to provide certified safety bounds under environmental perturbations~\cite{zeng2023safety}.
Recent research has pivoted toward making these models deployment-ready through techniques like Monte Carlo dropout (MC dropout), which approximates Bayesian inference by randomly masking neurons during inference~\cite{gal2016dropout}.
MC dropout is computationally simpler than Gaussian VI, and it can be enabled on most models without re-training, but it lacks the predictive depth and expressiveness that Gaussian VI provides for autonomous UAS operations~\cite{wang2025uncertainty}.
Consequently, current research is focused on hardware-software co-design: moving away from general-purpose GPUs toward specialized CMOS or emerging memory architectures designed explicitly for stochastic sampling.}

\subsection{BNN Hardware Acceleration}

\subsubsection{Digital Accelerators}

Digital accelerators must read distribution parameters, generate Gaussian samples via complex logic, and write them back to the weight array for processing.
Because GRNG and data movement dominate the energy budget, SOTA digital accelerators focus on GRNG optimization~\cite{thomas2013multiplierless, lee2005hardware} or pipeline/data-reuse techniques~\cite{fan2022accelerating, xu2021bayesian}.
Despite these efforts, the efficiency limits of digital GRNG and the frequency of memory operations required for BNN inference incur substantial overhead: for example, a system that generates and writes back a GRNG sample for each weight consumes approximately $6.2R\times$ more energy per \texttt{INT8} operation compared to a deterministic network~\cite{dorrance2023energy} (see Fig.~\ref{fig:overhead}).

\begin{figure}
    \centering
	\includegraphics[width=0.9\columnwidth]{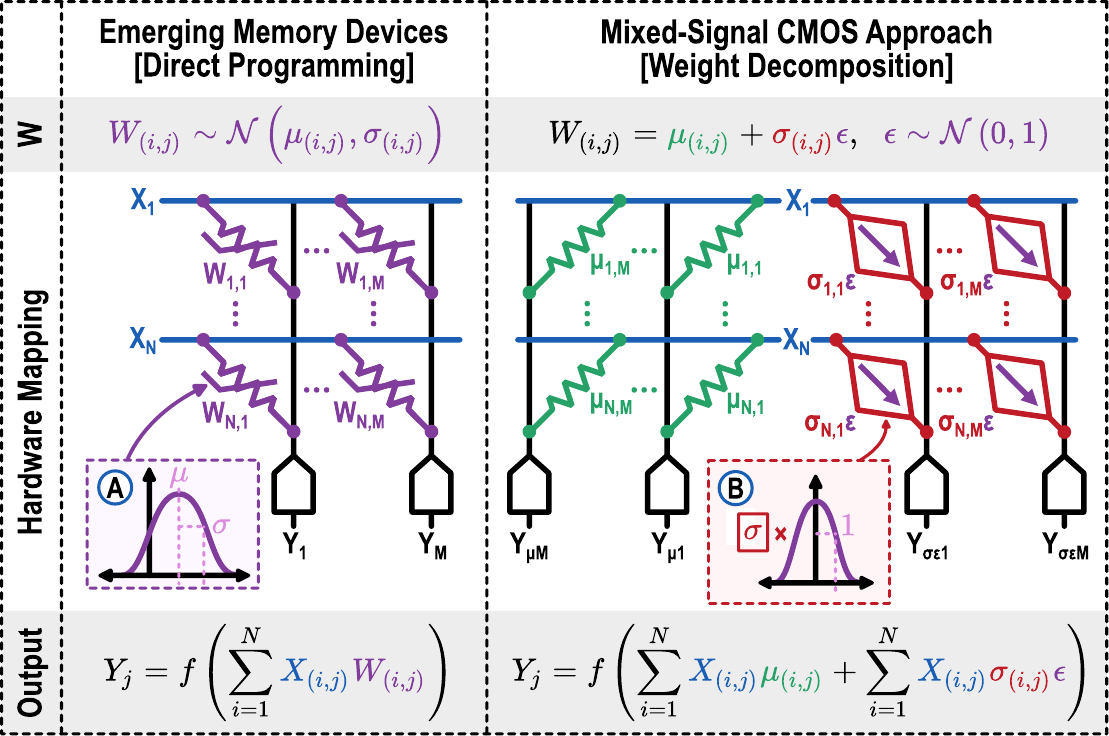}

    \caption{Comparison between approaches for mapping BNN stochastic computation to hardware.  \textbf{(A)} The single-device approach encodes $\mu$ and $\sigma$ in a single device.  \textbf{(B)} The weight decomposition approach computes $\mathbf{X} \cdot \mu$ and $\mathbf{X} \cdot \sigma\epsilon$ in separate arrays and sums their outputs via digital reduction logic.}
	\label{fig:mapping}

\end{figure}

\subsubsection{Single-Device GRNG}

In addition to offering nonvolatile storage, many emerging memory devices, such as phase-change memory, resistive RAM~\cite{lin2024high, bonnet2023bringing, malhotra2020exploiting}, or magnetic tunnel junctions~\cite{ahmed2023scalable, lu2022algorithm} exhibit stochastic behavior.
This stochasticity can be leveraged by programming the devices such that their conductance inherently represents the distribution parameters $\mu$ and $\sigma$, as shown in Fig.~\ref{fig:mapping}\textbf{(A)}.
While this method increases memory density by using a single device to encode the entire weight distribution, it introduces significant challenges.
First, the encoded value of $\sigma$ is often strongly correlated with $\mu$~\cite{bonnet2023bringing}.
Even if $\mu$ and $\sigma$ can be mathematically decoupled, their relative precisions are fixed by device physics, and BNNs require significantly less precision for $\sigma$ than for $\mu$~\cite{pei2024towards}.
Additionally, depending on the read scheme, sampling the device may also degrade the stored state~\cite{mulaosmanovic2020impact, shim2020investigation}, decreasing the accuracy of stored distributions over time.
Most significantly, this method limits technology scalability because programming precise distributions often necessitates large devices~\cite{kazemi2021fefet, sharifi2021application}, negating the benefits of advanced process nodes.
The relationship between programming precision and device size for FeFETs specifically is explored in more depth in Sec.~\ref{sec:single}.

\subsubsection{Weight Decomposition}
\label{sec:decomposition}

Weight decomposition, introduced in SOTA mixed-signal BNN accelerators~\cite{pei2024towards, liu2025va}, fully decouples $\mu$ and $\sigma$ by storing them in distinct subarrays.
The mean parameter is static, so it only needs to be processed once per input by the $\mu$ subarray \hl{to evaluate the deterministic portion of the layer.
The $\sigma\epsilon$ subarray performs the stochastic computation by embedding a localized GRNG per word, as shown in Fig.~\ref{fig:mapping}\textbf{(B)}.
When matrix-vector multiplication occurs, the in-word GRNG samples from a standard normal distribution $(\epsilon \sim \mathcal{N}(0,1))$, and the cell scales this dynamic sample by the cell's stored standard deviation $(\sigma)$ parameter.
After both subarrays process the shared input, their outputs are summed via digital reduction logic; the result is mathematically equivalent to multipling and accumulating samples from a parameterized Gaussian distribution, $\mathbf{W} \sim \mathcal{N}(\mu, \sigma^2)$.}
This approach relieves the pressure of representing a complex Gaussian distribution within a single device, and it could retain the density benefits of EMDs by storing each parameter bit in a single nonvolatile cell rather than a bulky 6T or 8T SRAM cell.
EMD-based GRNGs, such as the FeFET-based implementation detailed in Sec.~\ref{sec:clt}, also occupy less area, consume less energy, and are more robust to process variation \hl{than dynamic noise-based CMOS GRNGs.}

\begin{figure}
	\centering
	\includegraphics[width=0.95\columnwidth]{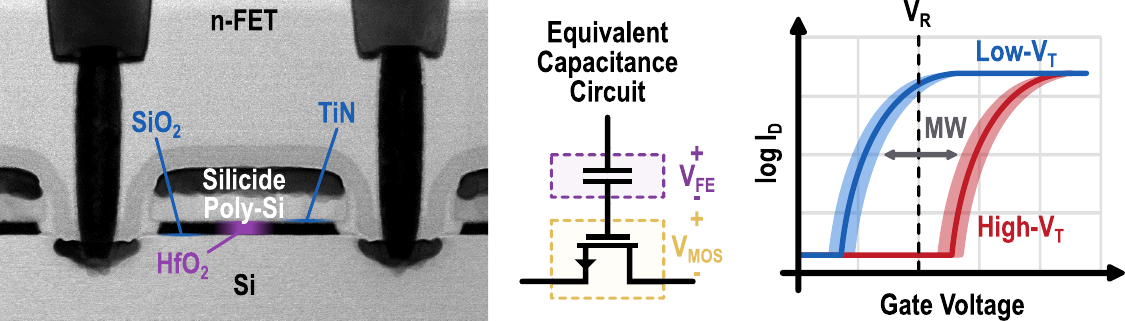}

    \caption{\hl{Cross-section of FeFET structure obtained with transmission electron microscopy depicting a} ferroelectric \ce{HfO2} layer integrated into the MOSFET gate stack.  The equivalent circuit models the series capacitance of the ferroelectric and gate dielectrics, and \hl{a representative $I_D$-$V_G$ transfer curve depicts how the low-$V_t$ and high-$V_t$ states create a memory window} (MW).} \label{fig:fedevice}

\end{figure}

\begin{figure}
    \centering
	\includegraphics[width=0.9\columnwidth]{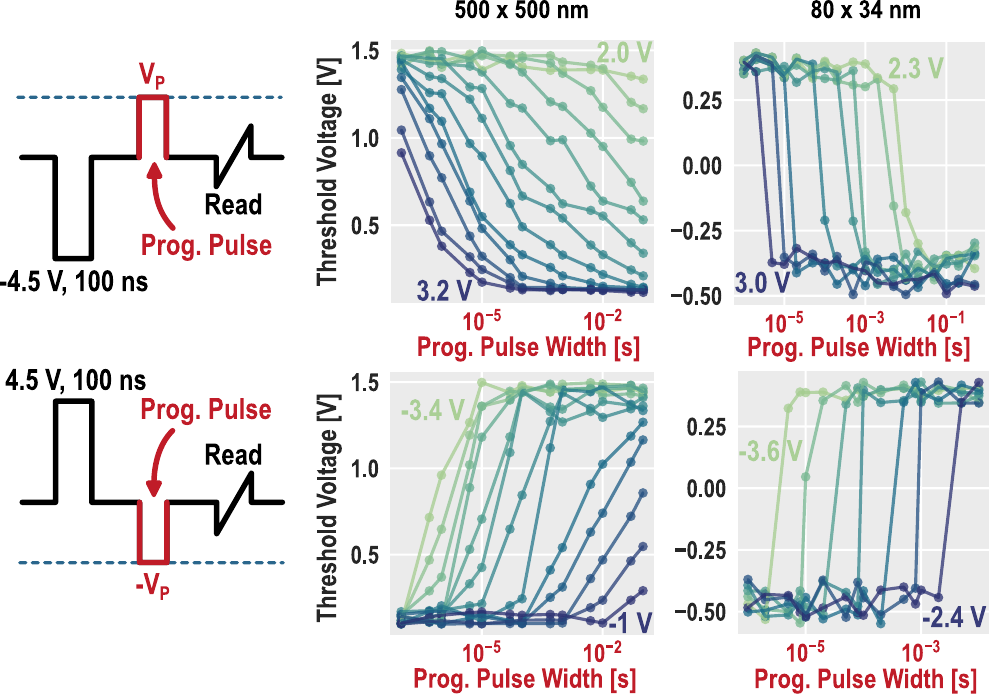}

	\caption{Measured FeFET switching mechanics from fabricated devices. Large devices provide fine control over intermediate states, making them essential for applications requiring precise programming.  In contrast, small FeFETs switch much more abruptly between distinct high-$V_t$ and low-$V_t$ states.}
    \label{fig:fe_programming}

\end{figure}

\begin{figure*}[ht]
    \centering
    \includegraphics[width=0.9\textwidth]{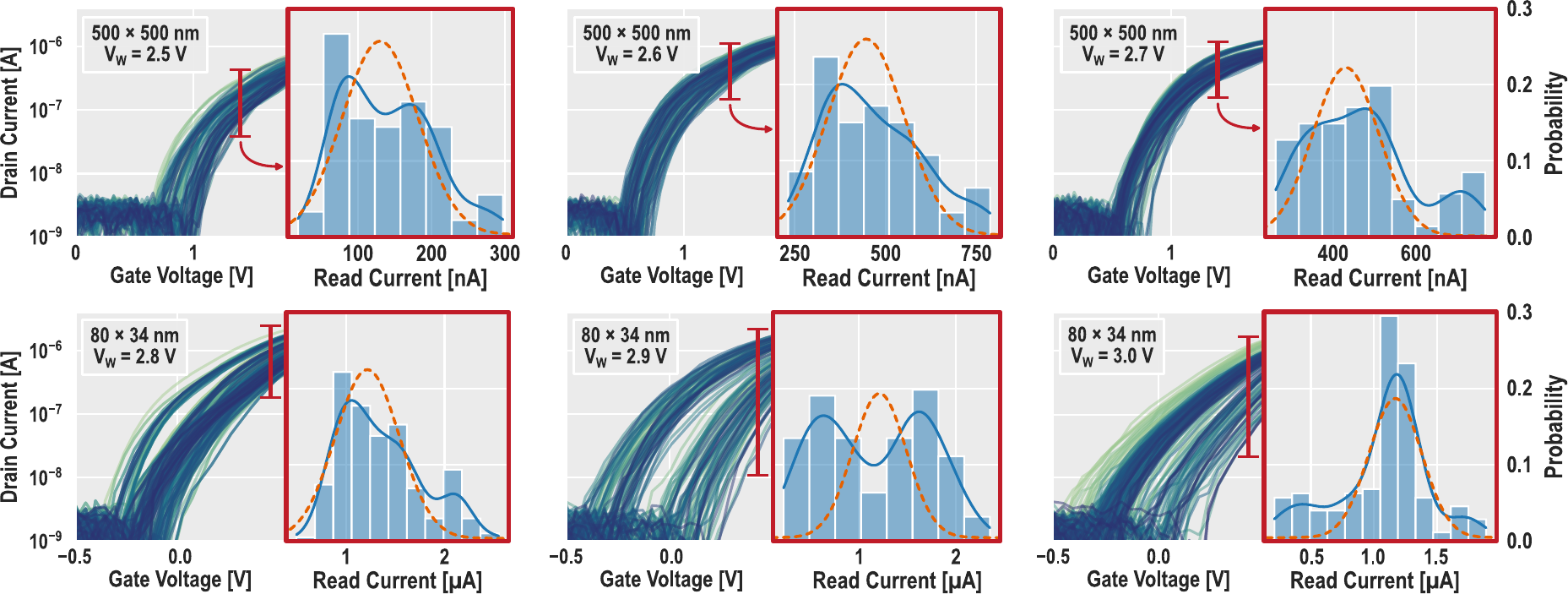}

    \caption{Measured $I_D$-$V_G$ curves for large and small fabricated FeFETs after programming with various write voltages. A large FeFET can approximate a Gaussian distribution (dotted orange line), but the abrupt switching behavior of small FeFETs results in high stochasticity and sensitivity to changes in programming pulse, making them unsuited for single-device GRNG.} \label{fig:single}

\end{figure*}

\subsection{Ferroelectric Field-Effect Transistors}

Though all of the emerging memory devices discussed in Sec.~\ref{sec:single} exhibit stochastic behavior, this work focuses on FeFETs due to their low read current and scalability~\cite{kirtania2024amorphous, manna2024variation, dunkel2017fefet}.
As illustrated in Fig.~\ref{fig:fedevice}, FeFETs integrate a ferroelectric layer (doped \ce{HfO2} for the devices measured in this work) into the gate stack of a standard MOSFET~\cite{ni2018critical}, \hl{which adds a ferroelectric capacitance in series with the gate dielectric capacitance.}
The polarization state of the ferroelectric material modulates the threshold voltage ($V_t$) of the FeFET: an upwards polarization leads to accumulation (high-$V_t$ state), while a downwards polarization induces an inversion layer in the channel (low-$V_t$ state).

This polarization persists after the removal of an external electric field, \hl{creating the nonvolatile memory window (MW) of Fig.~\ref{fig:fedevice}~\cite{qin2024understanding}.}
While multi-level programming is possible, it typically requires complex write schemes and restricts device scaling~\cite{kazemi2021fefet}.
Consequently, this work leverages binary, minimum-sized FeFETs to maximize density and scalability, relying on the collective statistics of the array rather than individual device precision.
The impact of FeFET sizing on programming characteristics is analyzed further in Sec.~\ref{sec:single}.
FeFETs have also been used to implement a cryptographically secure RNG~\cite{mulaosmanovic2017random}, but this approach requires rewriting the polarization state between samples, which severely limits device endurance as discussed in Sec.~\ref{sec:clt}.

%% file: 02_GRNG.tex
\section{GRNG Circuit Design}

\subsection{Single-FeFET GRNG}
\label{sec:single}

With careful tuning of pulse width and amplitude, a single FeFET can ideally be programmed to represent a continuum of conductance states.
However, the fidelity of this programming degrades as device geometries shrink.
Fig.~\ref{fig:fe_programming} demonstrates this scaling challenge: while a large device ($W$$\times$$L$ = $500$$\times$$500$\,\unit{\nm}) can be programmed to intermediate states, scaled FeFETs ($W$$\times$$L$ = $80$$\times$$34$\,\unit{\nm}) exhibit abrupt switching between high-$V_t$ and low-$V_t$ states~\cite{mulaosmanovic2017switching}.
Consequently, while \hl{the large devices measured in Fig.~\ref{fig:single} (top) can produce an approximately unimodal Gaussian distribution, the small devices in Fig.~\ref{fig:single} (bottom) yield bimodal or unpredictable distributions} unsuitable for parameter encoding.
Furthermore, single-device implementations lack robustness; for both device sizes, a mere \qty{100}{\milli\volt} deviation in programming voltage dramatically shifts the output distribution.
Finally, encoding both parameters in one device inherently couples the distribution's mean and standard deviation, restricting the representable space of Gaussian distributions and limiting model accuracy~\cite{bonnet2023bringing}.

\subsection{Central Limit Theorem-Based GRNG}
\label{sec:clt}

\begin{figure}
    \includegraphics[width=\columnwidth]{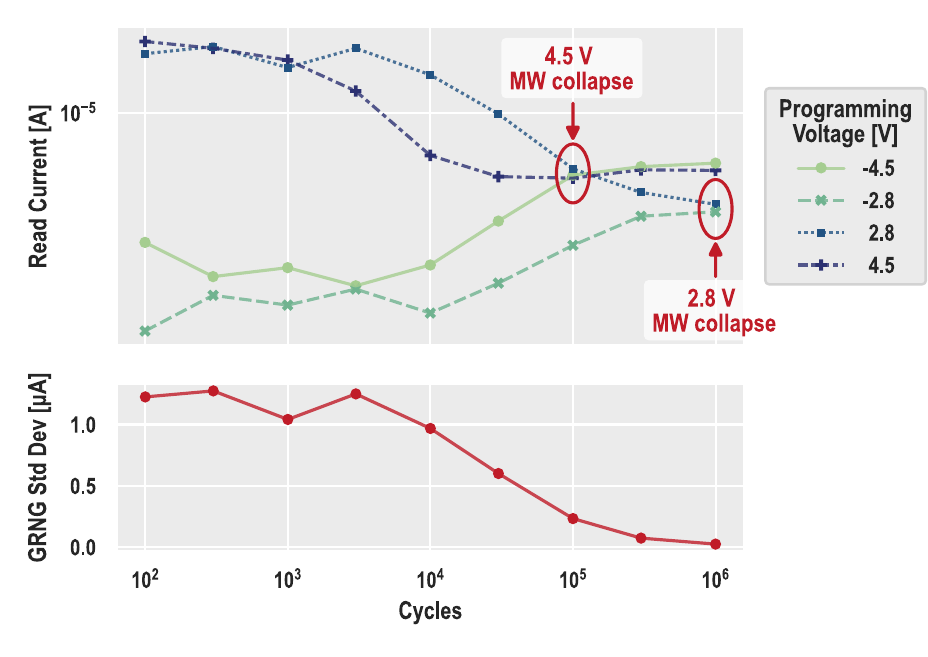}

    \caption{Measured FeFET endurance over time ($V_G = $\ \qty{1}{\volt}, $V_D = $\ \qty{0.05}{\volt}). While programming the device with a low-amplitude pulse (required for random state generation) delays the collapse of the memory window, it extends endurance by only one order of magnitude. Consequently, a CLT-GRNG relying on per-sample rewrites would suffer from collapsing output range until eventual failure.}
    \label{fig:endurance}

\end{figure}

To overcome the scalability limits of single-device programming, GRNGs may leverage the central limit theorem (CLT), which states that the sum of multiple independent random variables converges to a normal distribution regardless of the individual variables' distributions.
In this scheme, an array of minimum-size FeFETs is erased to a uniform state and then subjected to a low-amplitude programming pulse to induce random $V_t$ variation.
By accumulating the read currents of these devices onto a shared capacitor, the resulting voltage represents the sample mean and follows a Gaussian-like distribution.
This approach decouples the quality of the Gaussian distribution from the precision of individual devices, enabling the use of highly scaled, minimum-size FETs.

This CLT implementation would require rewriting the FeFET array before every read to generate a new random sample.
Device endurance is a known limitation of ferroelectrics~\cite{yuan2024spatial, agarwal2024study}.
\hl{As shown in Fig.~\ref{fig:endurance}, which plots FeFET read current over cycles with varying programming amplitudes,} using low-amplitude programming pulses extends endurance by approximately one order of magnitude.
\hl{However, Fig.~\ref{fig:endurance} also depicts the GRNG standard deviation subject to device wear, and our measurements indicate} the GRNG output range would collapse by \qty{50}{\percent} within 30,000 cycles.
In the context of continuous SAR inference, this lack of longevity is critical; even assuming a generous $10^{12}$ cycle endurance~\cite{kirtania2024amorphous}, a system operating at \qty{10}{\mega\hertz} (\qty{100}{\ns} FeFET write time) would fail within 30 hours.
Therefore, the GRNG must be designed to generate dynamic samples without reprogramming.

\begin{figure}
    \centering
	\includegraphics[width=0.9\columnwidth]{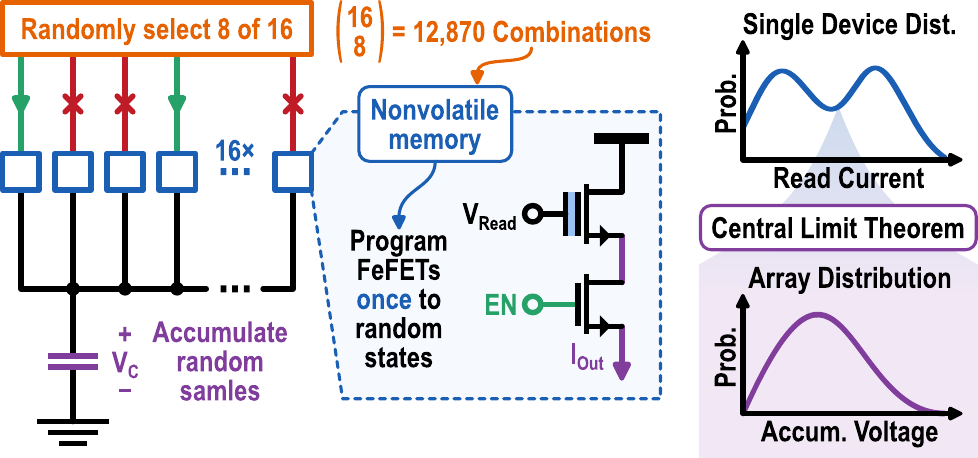}

	\caption{CLT-GRNG operating principles.  An array of 16 FeFETs is initially programmed to random states using a single low-amplitude pulse. For each GRNG sample, a selection circuit randomly enables 8 of the 16 devices, and their currents are accumulated on a sampling capacitor. Though individual device distributions may not be Gaussian---especially in highly-scaled nodes---their sum follows an approximately Gaussian distribution.}
    \label{fig:clt}

\end{figure}

\begin{figure}
    \centering
	\includegraphics[width=\columnwidth]{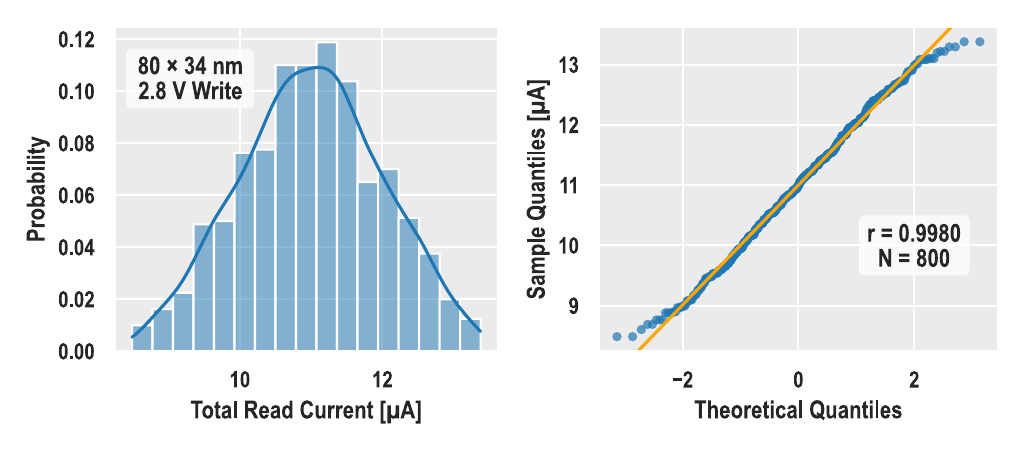}

    \caption{Representative CLT-GRNG output distribution using 16 $80$$\times$$34$\,\unit{\nm} FeFETs programmed at \qty{2.8}{\volt}.  Eight FeFETs are selected per the selection circuitry of Fig.~\ref{fig:swappers} during each sample.  The associated Q-Q plot shows minor deviation from an ideal Gaussian distribution (orange line).}
    \label{fig:clt_hist}

\end{figure}

\begin{figure}
	\centering
	\includegraphics[width=0.9\columnwidth]{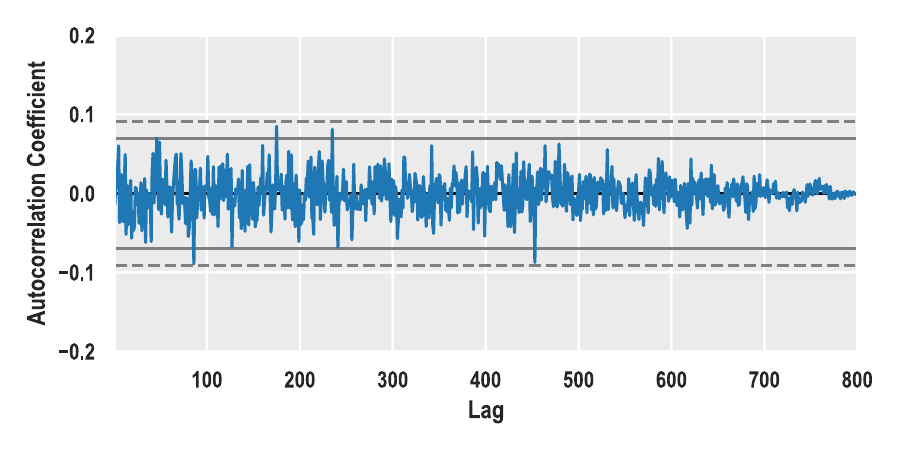}

	\caption{\hl{Autocorrelation of the CLT-GRNG output, with a Pearson correlation coefficient of $-0.0139$ at lag $1$ ($p=0.694$).}}
	\label{fig:correlation}
\end{figure}

In this work's \hl{in-word} CLT-GRNG, an array of sixteen FeFETs are programmed once to random $V_t$ states, as shown in Fig.~\ref{fig:clt}.
Rather than rewriting the devices, a digital selector circuit randomly enables eight of the sixteen FeFETs during each cycle, and the current from this subset is accumulated onto a \qty{1}{\femto\farad} capacitor.
This combinational approach yields $\binom{16}{8} = 12\,870$ unique current sums, ensuring \hl{the output distribution (shown as a histogram and normal probability plot in Fig.~\ref{fig:clt_hist})} retains a Gaussian shape without device degradation.
Although the output distribution is not statistically perfect (failing D'Agostino's $\mathrm{K^2}$ and Anderson-Darling tests) \hl{or cryptographically robust (failing the NIST SP 800-22 test suite)}, \hl{the output correlates strongly with an ideal Gaussian distirbution; the mean Q-Q correlation coefficient $r$ across 200 CLT-GRNG arrays is $0.9950$ (minimum $0.9667$).}
\hl{While the shape of the distribution is important, temporal independence is equally critical to avoid unwanted predictability.
As shown in Fig.~\ref{fig:correlation}, the Pearson autocorrelation coefficient at lag 1 is $-0.0139$ ($p=0.694$), suggesting that the consecutive samples are effectively uncorrelated~\cite{box2015time}.}
Furthermore, BNNs are inherently resilient to imperfect distributions~\cite{pei2024towards}, and the subsequent multiplication with a quantized $\sigma$ effectively masks minor distribution defects.
The impact of this output distribution on SAR accuracy and uncertainty quantification (UQ) is rigorously evaluated in Sec.~\ref{sec:sar}.

\begin{figure}
    \centering
    \includegraphics[width=0.6\columnwidth]{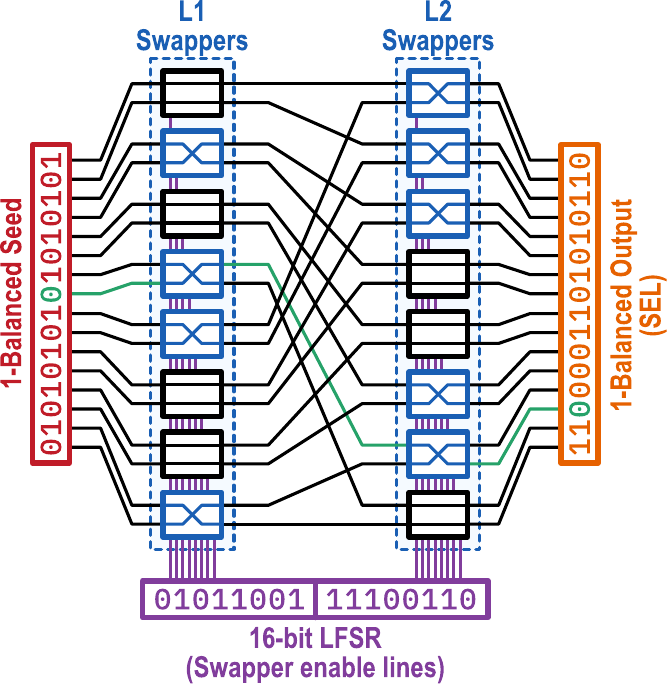}

    \caption{Digital circuit design for random FeFET selection. A 16-bit LFSR randomly drives two layers of swappers. A fixed input vector containing eight \texttt{1}s and eight \texttt{0}s is permuted by the swappers, guaranteeing that exactly 8 of the 16 FeFETs are enabled per cycle. The selection lines (output) are shared across all CLT-GRNG cells in the macro.}
    \label{fig:swappers}
\end{figure}

\begin{figure*}
    \centering
	\includegraphics[width=0.9\textwidth]{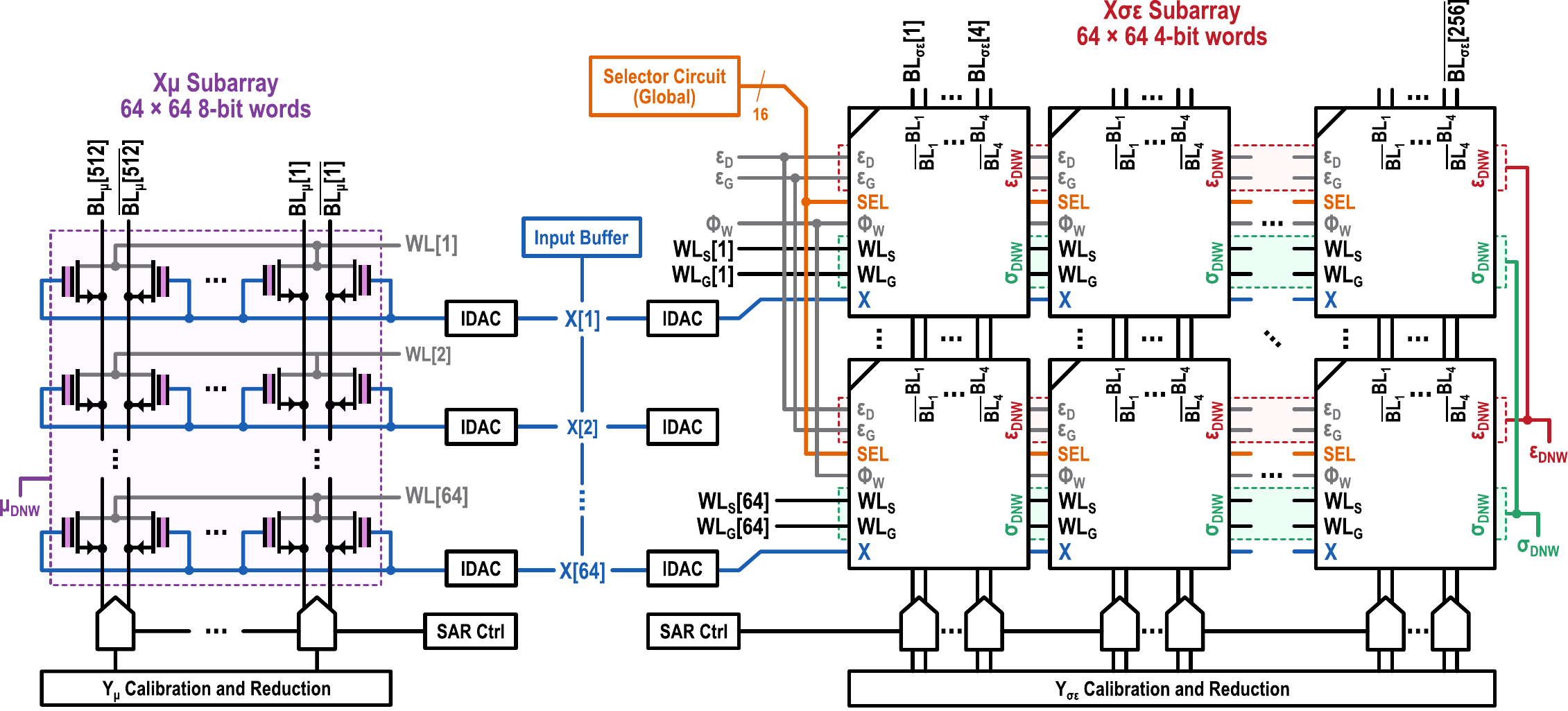}

	\caption{CIM tile block diagram.  Two 64$\times$64 subarrays implement the mean ($\mu$) and variance-scaled sampling ($\sigma\epsilon$) separately. Shared control signals and selection lines between CLT-GRNGs minimize peripheral circuitry. The $\mathbf{X}\mu$ subarray can execute concurrently with $\mathbf{X}\sigma\epsilon$, or they may be operated independently.}
	\label{fig:clt_arch}

\end{figure*}

The selection circuitry is purely digital, consisting of two layers of wire swappers (see Fig.~\ref{fig:swappers}).
The first layer randomly swaps adjacent bits (e.g., bit \texttt{0} and \texttt{1}), while the second layer swaps the $\mathrm{n^{th}}$ bit with the $\mathrm{\left(n+8\right)^{th}}$ bit.
A 16-bit LFSR drives the swap control lines; the first 8 bits control the first layer, and the remaining 8 bits control the second.
By feeding a fixed input vector with exactly eight \texttt{1}s into this shuffling network, the circuit guarantees that exactly eight FeFETs are selected every cycle.
Because the individual FeFETs in the array are pre-programmed to random states, this selection sequence is uncorrelated with the device values.
This allows the selection control lines to be shared across all CLT-GRNG cells in the CIM tile, significantly amortizing the area and power overhead of the digital logic.

\subsubsection{Static Offset Compensation}

The CLT-GRNG produces a current sum with \hl{an offset (non-zero) mean, which is affected by operating temperature. Additionally, programming instability, process variation, and device mismatch cause every GRNG instance to exhibit a unique---but static---offset} ($\Delta\epsilon_{\left(i, j\right)}$).
Without compensation, $\Delta\epsilon_{\left(i, j\right)}$ distorts the stored weights per Eq.~\ref{eq:badweight}:

\begin{equation}
    \label{eq:badweight}
    w_{\left(i, j\right)} = \mu_{\left(i, j\right)} + \sigma_{\left(i, j\right)} \left( \epsilon + \Delta\epsilon_{\left(i, j\right)} \right)
\end{equation}

We mitigate this via one-time weight offsetting.
Since the mean offset is static, we measure the deviation $\Delta\epsilon_{\left(i, j\right)}$ and adjust the stored mean parameter $\mu$ accordingly:

\begin{gather}
    w_{\left(i, j\right)} = \mu'_{\left(i, j\right)} + \sigma_{\left(i, j\right)} \epsilon \\
    \mu'_{\left(i, j\right)} = \mu_{\left(i, j\right)} - \sigma_{\left(i, j\right)}\Delta\epsilon_{\left(i, j\right)}
\end{gather}

\hl{Variations in CLT-GRNG output standard deviation, which are also static, can be similarly compensated by adjusting the stored $\sigma$ value.
For example, if the standard deviation of instance $\left(i, j\right)$ is scaled by $\alpha_{\left(i, j\right)}$, then the stored $\sigma$ should be updated accordingly:}

\begin{equation}
    \sigma'_{\left(i, j\right)} = \frac{\sigma_{\left(i, j\right)}}{\alpha_{\left(i, j\right)}}
\end{equation}

The compensation process consumes $54 + 458N$\,\unit{\pico\joule} and takes $12.8 + 0.64N$\,\unit{\micro\second}, where $N$ is the number of samples used to estimate the mean.
We note that this compensation consumes a portion of the dynamic range of the $\mu$ subarray.
Given the measured distribution in Fig.~\ref{fig:clt_hist} (mean: \qty{10.1}{\micro\ampere}, SD: \qty{0.993}{\micro\ampere}) and a 4-bit $\sigma$ representation, the correction term $\sigma\Delta\epsilon_{\left(i, j\right)}$ ranges up to 162.72.
This reduces the effective precision of $\mu$ by approximately 1.5 bits (to 6.54 bits), a trade-off that is acceptable given the significant energy gains of the write-free architecture.

%% file: 03_Tile.tex
\section{CIM Tile Architecture}
\label{sec:tile}

\hl{Fig.~\ref{fig:clt_arch} provides the CIM tile architecture, which adapts the weight decomposition method of Sec.~\ref{sec:decomposition} to FeFET GRNG and weight storage.
It consists of two 64$\times$64 subarrays:} a $\mathbf{X}\cdot\mu$ subarray (``$\mu$ subarray'') storing static 8-bit weights, and a $\mathbf{X}\cdot\sigma\epsilon$ subarray (``$\sigma\epsilon$ subarray'') storing 4-bit standard deviations with CLT-GRNGs \hl{embedded into each $\sigma\epsilon$ word}.
This mixed-precision architecture leverages the observation that model accuracy is primarily dictated by $\mu$ precision, while UQ performance depends on $\sigma$ precision~\cite{pei2024towards}.
Both subarrays operate at \qty{100}{\mega\hertz} and can execute concurrently or independently.
As detailed in Sec.~\ref{sec:clt}, the random selector circuit is global; its 16-bit output bus is buffered and distributed to every $\sigma\epsilon$ cell in the tile.
The input vector ($\mathbf{X}$) is fed to the wordlines (WLs) via current digital-to-analog converters (IDACs), which drive the WLs with analog voltages such that the bitcell current is linear with the input code.
During operation, the differential bitlines ($BL_j$ and $\overline{BL_j}$) are precharged to \qty{0.8}{\volt} and subsequently discharged by the active bitcells.
The resulting BL voltages are digitized by successive approximation register analog-to-digital converters (ADCs).
By dedicating a pitch-matched, 6-bit ADC to each column, this architecture eliminates column multiplexing and enables full-tile, single-cycle matrix-vector multiplication.

\subsection{$\mathbf{X}\sigma\epsilon$ Multiply-Accumulate Cell}

\begin{figure}
    \centering
    \includegraphics[width=\columnwidth]{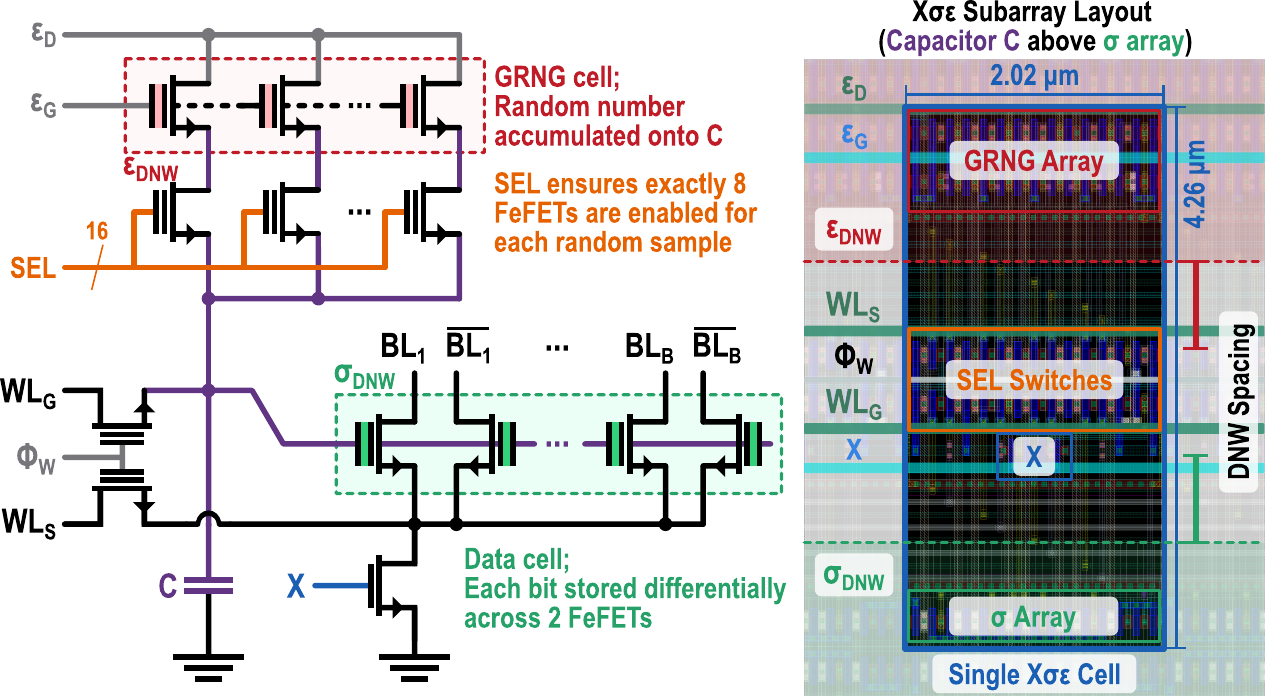}

    \caption{$\sigma\epsilon$ MAC cell schematic.  The CLT-GRNG array and $\sigma$ storage occupy separate n-wells to support independent erase/program cycles.  Inset: physical layout of the $X\sigma\epsilon$ MAC cell within the subarray.  \hl{The capacitor $C$ is a metal fringe capacitor above the $\sigma$-storing FeFETs.}}
	\label{fig:clt_xse}

\end{figure}

While the $\mu$ subarray consists of standard differential FeFET bitcells, the $\sigma\epsilon$ subarray utilizes a more complex structure to support in-memory stochastic computation.
\hl{Fig.~\ref{fig:clt_xse} provides the $\sigma\epsilon$ multiply-accumulate (MAC) cell schematic,} which integrates a CLT-GRNG (producing the random variable $\epsilon$) and a differential FeFET bank storing the standard deviation $\sigma_{\left(i, j\right)}$.
Operation proceeds in three phases: First, the CLT-GRNG accumulates current from a randomly selected subset of FeFETs onto the sampling capacitor $C$ while the BLs are precharged to $V_{DD}$.
Second, the voltage on $C$ gates the $\sigma$ bitcells, modulating their conductance to represent the product $\epsilon \cdot \sigma_{\left(i, j\right)}$.
Finally, the input IDACs drive the WLs, gating the current flow to compute the final term $\mathbf{X} \cdot \sigma\epsilon$.

All FeFETs reside in deep n-wells to allow for body biasing (see Sec.~\ref{sec:negvoltage}), \hl{which are visibile as shaded regions in the layout of Fig.~\ref{fig:clt_xse}.}
Assuming a weight-stationary dataflow where the entire model fits on-chip, $\sigma_{\left(i, j\right)}$ is written only once.
However, to support potential updates, the $\sigma$ storage resides in a separate, isolated n-well from the CLT array.
This isolation protects the CLT-GRNG's random states during $\sigma$ reprogramming.
While the PDK-mandated spacing between deep n-wells accounts for \qty{42}{\percent} of the $\sigma\epsilon$ cell area, the separate well---and thus some of the spacing overhead---can be removed if independent erasure of the $\sigma$ and CLT-GRNG cells is not required by the application.
The CLT capacitor $C$ is implemented as a mental fringe capacitor above the cell to ensure matching~\cite{tripathi2014mismatch}.
All of the non-ferroelectric transistors within the $\sigma\epsilon$ cell, including the disconnect switches, the CLT selection switches, and the BL discharge transistor, are thick-oxide devices.
These devices are physically larger than standard transistors \hl{(reflected in the area breakdown of Sec.~\ref{sec:area})}, but they withstand the high voltages required for FeFET programming without risking oxide breakdown.

\subsection{Erasing and Programming Cells}
\label{sec:negvoltage}

\begin{figure}
	\centering
	\includegraphics[width=0.90\columnwidth]{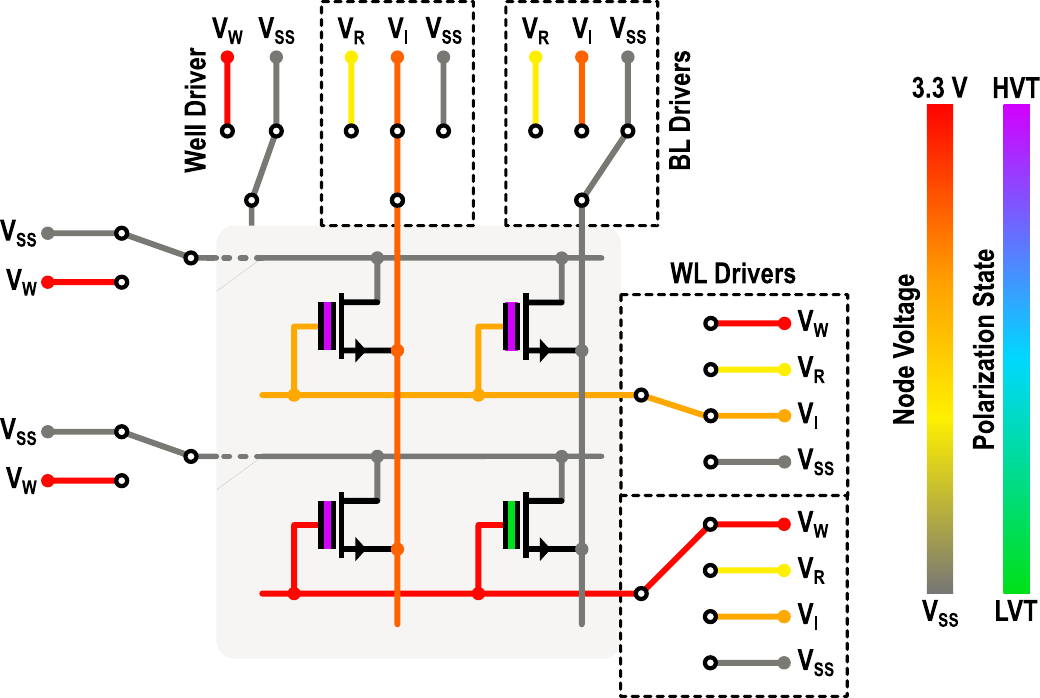}

    \caption{FeFET array write scheme.  A high write voltage programs the target cell (bottom left) while wordline (WL) and bitline (BL) inhibit voltages suppress inadvertent programming of adjacent cells.  The isolated n-well supports bulk erase via body bias, eliminating the need for negative gate voltages.}
	\label{fig:feoperation}

\end{figure}

Traditional FeFET designs often utilize negative gate voltages to reset ferroelectric polarization~\cite{yin2020fecam}.
However, generating negative voltages risks forward-biasing substrate junctions, adds level-shifter circuitry, and introduces more peripheral circuit spacing between buried wells.
The proposed design eliminates negative voltages entirely by employing a bulk erase strategy: a \hl{high voltage is applied to the body (n-well), drain, and source of each FeFETs while their gates are grounded} to reset all FeFETs in the array \hl{simultaneously~\cite{vardar202528nm, yin2024ferroelectric}.}
\hl{This erase scheme is similar to the drain-erase scheme~\cite{wang2020drain1}, but it provides additional control over the channel conditions.}
After reset, cells are programmed row-wise.
Although this approach requires additional array well contacts, it significantly reduces the area and power overhead of the peripheral circuitry.

Parasitic programming (``write disturb'') is a critical concern in crossbar arrays.
Since the polarization state is determined by the potential difference between the gate and the channel, applying a write voltage to a target cell can inadvertently disturb other devices sharing the same wordline or bitline.
To mitigate this, we employ an inhibit scheme (see Fig.~\ref{fig:feoperation}) where inhibit voltages are applied to unselected WLs and BLs.
This reduces the effective gate-channel voltage for non-target devices, confining polarization switching to the selected cell.
During writes, the drain is \hl{grounded through a large resistance to reduce} unwanted leakage current.

\begin{figure}
	\centering
	\includegraphics[width=\columnwidth]{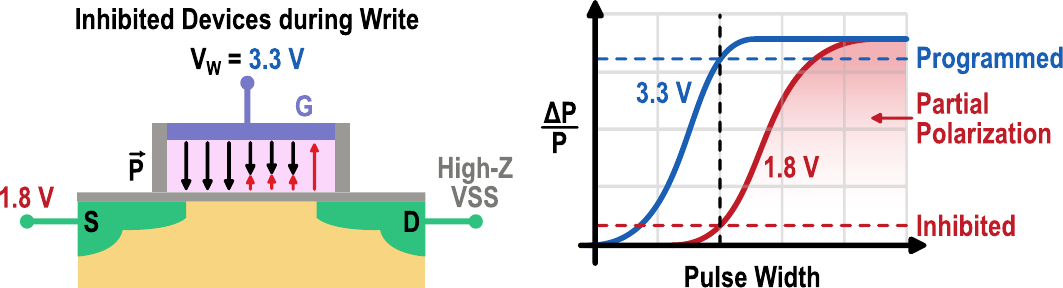}

    \caption{Partial polarization of inhibited cells while programming.  \hl{After fabrication, the optimal programming pulse width can be identified by ploting normalized polarization change $\left(\frac{\Delta P}{P}\right)$ versus pulse width. The optimal pulse (dotted black line) will fully switch the target cell while fully inhibiting partial polarization in non-target cells.}}
	\label{fig:partialpolar}

\end{figure}

Given a sufficiently long programming pulse, even inhibit voltages can induce partial or full polarization in non-target cells (Fig.~\ref{fig:partialpolar}).
To suppress this, we can tune the programming pulse width such that the target cell achieves full polarization while the inhibited cells---subjected to a lower effective voltage---remain in their original state.
\hl{Fig.~\ref{fig:partialpolar} also provides a representative characterization of the optimal programming pulse, which must be identified on a per-die basis due to fabrication inconsistencies.}
For the CLT-GRNG specifically, full polarization swing is not required; the programming pulse is calibrated to roughly \qty{2.8}{\volt} to produce a high-entropy mix of high-$V_t$ and low-$V_t$ states.

%% file: 04_Evaluation.tex
\section{Hardware Evaluation}

\begin{table*}
    \centering
    \caption{CLT-GRNG Accelerator Compared to Other BNN Accelerators}
    \label{tab:clt}

    \begin{tabular*}{\textwidth}{@{\extracolsep{\fill}} r|cccccc }
        \toprule

        \multicolumn{1}{l}{} & \textbf{This Work} &
        \textbf{\cite{liu2025va}} &
        \textbf{\cite{dorrance2023energy}} &
        \textbf{\cite{you2025sttbnn}} &
        \textbf{\cite{lu2022algorithm}} &
        \textbf{\cite{xu2021bayesian}} \\

        \midrule

        \textbf{BNN Implementation} &
        \begin{tabular}[c]{@{}c@{}}Gaussian\\ Weights\end{tabular} &
        \begin{tabular}[c]{@{}c@{}}Gaussian\\ Weights\end{tabular} &
        \begin{tabular}[c]{@{}c@{}}Gaussian Output\\ Perturbation\end{tabular} &
        \begin{tabular}[c]{@{}c@{}}Single-Sample\\ Gaussian Weights\end{tabular} &
        \begin{tabular}[c]{@{}c@{}}Gaussian Output\\ Perturbation\end{tabular} &
        \begin{tabular}[c]{@{}c@{}}Gaussian Output\\ Perturbation\end{tabular} \\

        \textbf{Uncertainty Estimation?} &
        \textbf{\textcolor{green5}{Yes}} &
        \textbf{\textcolor{green5}{Yes}} &
        \textbf{\textcolor{green5}{Yes}} &
        \textbf{\textcolor{red4}{No}} &
        \textbf{\textcolor{green5}{Yes}} &
        \textbf{\textcolor{green5}{Yes}} \\

        \textbf{RNG Type} &
        CLT &
        Thermal &
        TI-Hadamard &
        CLT &
        MRAM Bitstream &
        Box-Muller \\

        \textbf{Technology} {[\unit{\nm}]} &
        \begin{tabular}[c]{@{}c@{}}22\\ (FeFET)\end{tabular} &
        \begin{tabular}[c]{@{}c@{}}65\\ (CMOS)\end{tabular} &
        \begin{tabular}[c]{@{}c@{}}22\\ (CMOS)\end{tabular} &
        \begin{tabular}[c]{@{}c@{}}22\\ (STT-MRAM)\end{tabular} &
        \begin{tabular}[c]{@{}c@{}}22\\ (SOT)\end{tabular} &
        \begin{tabular}[c]{@{}c@{}}16\\ (FPGA)\end{tabular} \\

        \textbf{Macro Area} {[\unit{\mm\squared}]} &
        0.0964 &
        0.106 &
        2.25 &
        3.49 &
        --- &
        --- \\

        \textbf{GRNG Area} {[\unit{\micro\meter\squared}]} &
        5.11 &
        24.9 &
        3870$^\ast$ &
        --- &
        \textcolor{green5}{\textbf{0.017}$^\ast$} &
        --- \\

        \textbf{GRNG Eff.} {[\unit{\femto\joule\text{/Sa}}]} &
        \textcolor{green5}{\textbf{0.640}} &
        360 &
        1080 &
        --- &
        1474 &
        5400 \\

        \textbf{GRNG Tput.} {[\unit{\text{GSa/s}}]} &
        \textcolor{green5}{\textbf{40.96}} &
        5.12 &
        4.65--7.31 &
        --- &
        --- &
        8.88 \\

        \textbf{\hl{Tile Energy Eff.}} {[\unit{\text{TOPS/W}}]} &
        17.8 &
        2.02 &
        1.17 &
        \textcolor{green5}{\textbf{104.5}} &
        --- &
        --- \\

        \textbf{Compute Density} {[\unit{\text{TOPS/}\mm\squared}]} &
        \textcolor{green5}{\textbf{1.27}} &
        0.889 &
        0.515 &
        0.036 &
        --- &
        --- \\

        \bottomrule

    \end{tabular*}%

    {\raggedright $^\mathrm{\ast}$Estimated from reported results. \par}

\end{table*}

\subsection{Bayesian CIM Tile Performance}

System-level performance was \hl{extrapolated from measured data from fabricated FeFET devices on a test die, which provides the switching mechanics, programming characteristics, device endurance, and the CLT-GRNG distribution.
This data was supplemented with SPICE simulations using a calibrated Preisach 22\,nm FeFET model~\cite{ni2018circuit} for tile-level energy and latency estimates.
The SPICE simulations also included parasitic capacitance and resistance extracted from the array layout.}

\subsubsection{Energy and Latency}

Writing the $\mu$ subarray and $\sigma\epsilon$ subarrays consumes \qty{92.7}{\pico\joule} and \qty{46.3}{\pico\joule}, respectively, using a \qty{4.0}{\volt} write voltage.
During inference, a complete tile matrix-vector multiplication (MVM) consumes \qty{688}{\pico\joule} under worst-case switching conditions (all cells conducting).
The 6-bit SAR ADCs, operating at an efficiency of \qty{14}{\femto\joule}/conv-step\footnote{\qty{14}{\femto\joule}/conv-step represents a Pareto-optimal design for a 6-bit, \qty{100}{\mega\hertz} ADC~\cite{adc_survey}.}, dominate the power budget, accounting for \qty{99}{\percent} of the total read energy.
In contrast, the CLT-GRNG contributes only \qty{0.4}{\percent} of the total energy.
Because the random selection circuitry is global and shared across the tile, its per-cell energy cost is negligible (\qty{134}{\atto\joule} of the reported \qty{640}{\atto\joule} per sample), despite the block consuming a moderate \qty{550}{\femto\joule} per cycle globally.
When the $\sigma\epsilon$ subarray operates independently---as required for the multiple sampling iterations in the Bayesian layers---it consumes \qty{230}{\pico\joule} per MVM, with the GRNG contributing just \qty{0.7}{\percent} of the subarray's energy.

\subsubsection{Area Analysis}
\label{sec:area}

\begin{figure}
    \centering
    \includegraphics[width=0.8\columnwidth]{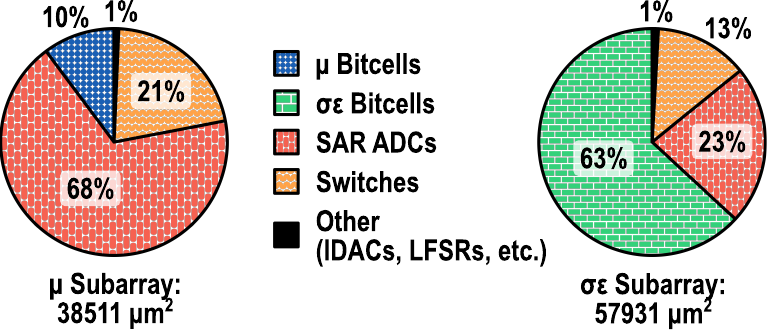}

    \caption{Area breakdown of the two CIM subarrays.  The $\mu$ bitcells consist of only two minimum-size FeFETs, resulting in an area dominated by the pitch-matched ADCs. The $\sigma\epsilon$ cells are larger due to the inclusion of the CLT-GRNG, $\sigma$ storage, and high-voltage switches.}
    \label{fig:clt_area}

\end{figure}

Layouts for all components (MAC cells, decoding logic, drivers, IDACs, and ADCs) were designed in a commercial 65\,nm PDK and scaled to a 22\,nm node by comparing the foundry's reported 65\,nm SRAM bitcell area to its 22\,nm SRAM bitcell.
\hl{Fig.~\ref{fig:clt_area} provides a breakdown of each subarray;} the combined CIM tile occupies \qty{0.0964}{\mm\squared}, and the $\sigma\epsilon$ subarray accounts for \qty{60.1}{\percent} of this total.
\hl{Each BL and WL driver must support multiple voltage levels, which requires additional switches or transmission gates.
In addition, the magnitudes of the write and inhibit voltages are greater than the oxide breakdown voltage for core FETs, necessitating thick-oxide switches.
These thick-oxide switches also require level shifters to boost the logic signals to fully actuate the switch.
In total, the driver circuits introduce significant area overheads compared to a CMOS solution, accounting for} \qty{21.3}{\percent}  and \qty{13.3}{\percent} \hl{of the $\mu$ and $\sigma\epsilon$ subarrays, respectively.}
Within the $\sigma\epsilon$ subarray, the PDK-mandated deep n-well spacing required to isolate the $\sigma$ storage from the CLT array imposes a significant overhead: the bitcells (which also include the CLT-GRNG array and additional switches) account for \qty{63.1}{\percent} of the subarray area.
In contrast, the $\mu$ cells, which are a common deep n-well, are highly dense, occupying only \qty{10.2}{\percent} of their respective subarray.
Despite the complexity of the CLT logic, the GRNG cells themselves contribute only \qty{36.1}{\percent} of the $\sigma\epsilon$ subarray area, validating the area efficiency of the shared-selector architecture.

\subsubsection{Comparison with State-of-the-Art}

Table~\ref{tab:clt} compares this work to other SOTA BNN accelerators.
Our write-free approach yields a $560\times$ improvement in GRNG efficiency compared to the most efficient reported BNN GRNG~\cite{liu2025va}, as FeFET reads consume nearly two orders of magnitude less energy than writes.
Furthermore, eliminating long write pulses allows our design to achieve a tile GRNG throughput of 40.96\,GSa/s, the highest among the compared works.
While \cite{you2025sttbnn} demonstrates high computation efficiency, its primary design goal was noise resilience, not uncertainty estimation; as such, it reuses samples generated by a sequential GRNG and thus does not support uncertainty estimation via probabilistic outputs without severely limiting system throughput.
Additionally, while \cite{lu2022algorithm} achieves higher GRNG density (300$\times$), it relies on SOT-MRAM bitstreams that consume over three orders of magnitude more energy per sample.

\subsection{Search and Rescue Detection Performance}
\label{sec:sar}

To evaluate SAR performance, we trained models on the Search and Rescue Dataset (SARD)~\cite{sambolek2021automatic}, which consists of aerial imagery captured at altitudes ranging from \qty{15}{\meter} to \qty{75}{\meter}.
This dataset presents several challenges: First, the size of the bounding boxes representing subjects varies depending on the recording height; at higher altitudes, the subjects appear smaller and occupy fewer pixels, making them significantly harder to detect.
Person detection within SARD is further complicated by occlusion and camouflage, as individuals are often sheltered by vegetation, hidden behind stones, or visually fused with the ground.
Finally, unlike urban datasets where people are typically standing or walking, this dataset features subjects in atypical postures such as lying down, kneeling, or sitting to simulate the behavior of injured or exhausted individuals.

\subsubsection{Model Architecture and Implementation}

We used YOLO26n~\cite{yolo26_ultralytics}, the smallest YOLO variant, as the baseline CNN.
The BNN implementation mimics the YOLO26n architecture but replaces the final 1D projection layer with Bayesian weights.
Converting only the last layer balances computational cost with UQ capability~\cite{pei2024towards}.
During inference, this final layer is sampled 20 times to generate an output distribution.
Given the macro size of 64$\times$64, the final projection layer requires 24 full Bayesian CIM tiles.
The preceding deterministic layers are processed using \texttt{im2col}~\cite{zhou2021characterizing} mapped to 1659 $\mu$-only subarrays to conserve area.
The total deployment requires \qty{76.0}{\mm\squared} of macro area with every weight stored on-chip to avoid FeFET writes.
Softmax layers and self-attention cannot be computed directly on the CIM tile because they require intermediate writes, so they would have to be computed on a separate scalar functional unit.
However, the softmax and self-attention layers correspond only to \qty{2.62}{\percent} of the model's total activations.
Excluding these layers, the end-to-end macro energy and latency are \qty{3.70}{\milli\joule} and \qty{13.8}{\milli\second} (\qty{72.2}{\text{FPS}}), respectively.
Down-sampling the video feed framerate for detection inferences can reduce total model power; for example, at \qty{24}{\text{FPS}}, the FeFET CIM macros consume \qty{88.7}{\milli\watt} of power.

\subsubsection{Accuracy and Uncertainty Quantification}

Both models were trained for up to 500 epochs using early stopping with a patience of 50 epochs.
The CNN converged in 225 epochs, and the BNN in 213.
To measure accuracy, we compare the mean average precision (mAP) with an intersection over the union (IOU) of \qty{50}{\percent}.
In other words, the bounding box drawn by the model most overlap with the ground truth bounding box by at least \qty{50}{\percent} in order to classify the prediction as correct.
The BNN and CNN demonstrate effectively equivalent detection accuracy (mAP-50 $0.8588$ vs. $0.8573$).
Crucially, using the imperfect CLT-GRNG output distribution does not result in accuracy loss ($0.8607$, or $+$\qty{0.2}{\percent}), \hl{though the effects of aging and read disturb were not included in this assessment}.

\begin{figure}
    \centering
	\includegraphics[width=\columnwidth]{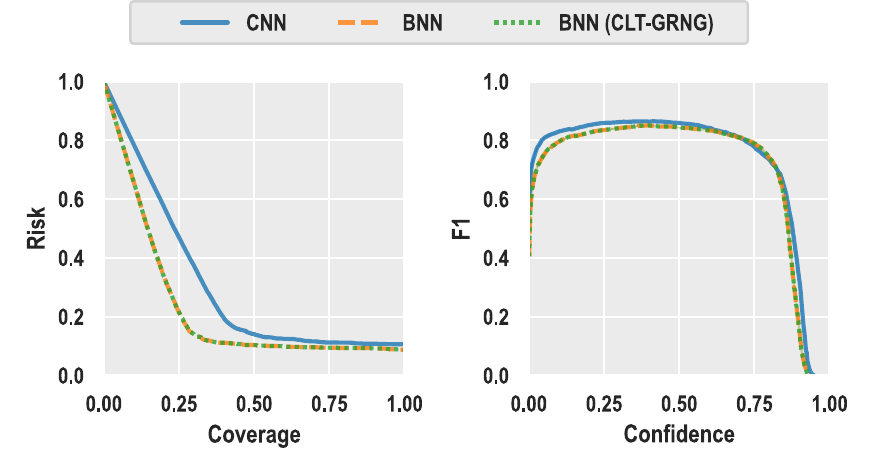}

	\caption{SARD accuracy and UQ performance comparison. The BNN reduces the risk associated with all coverage levels compared to the CNN, reducing the number of detections that must be physically verified. The CLT-GRNG closely tracks the ideal GRNG performance.}
    \label{fig:uq_sard}

\end{figure}

\begin{figure*}
    \centering
	\includegraphics[width=0.85\textwidth]{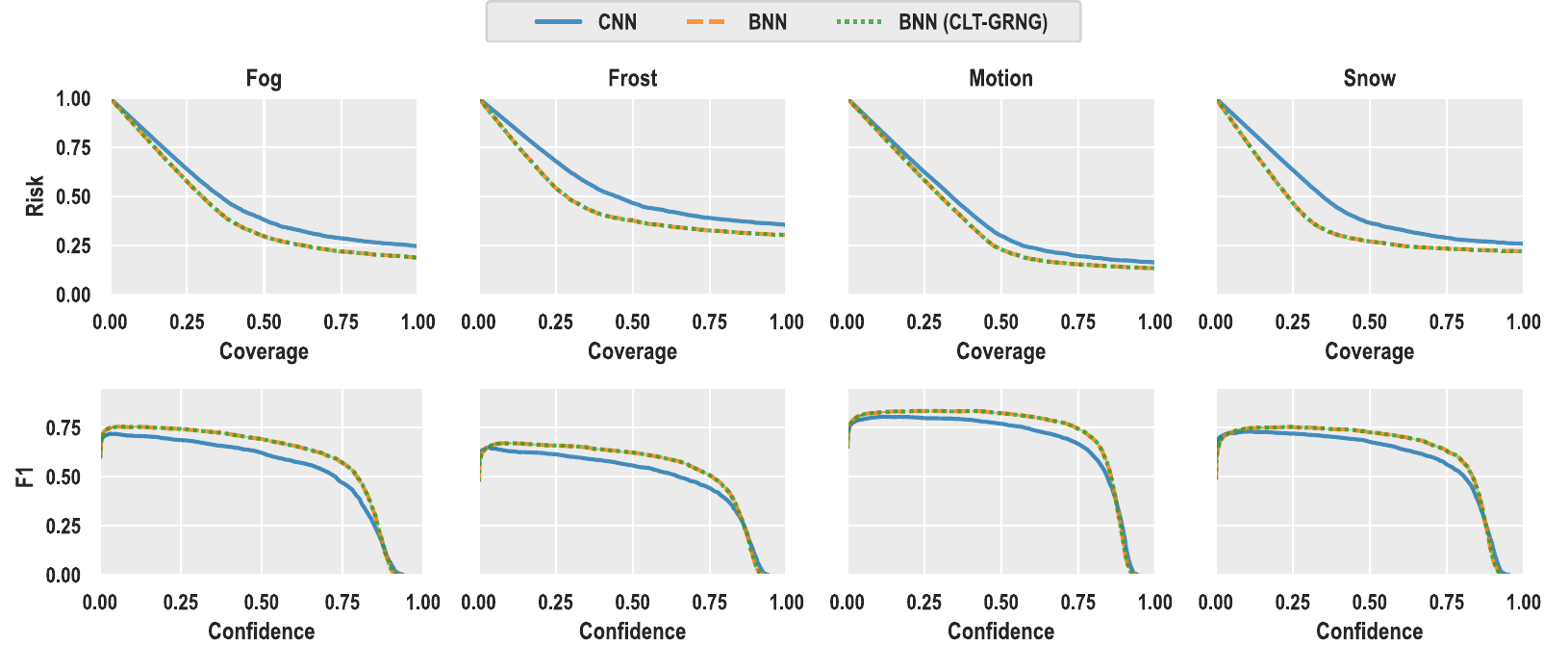}

	\caption{Accuracy and UQ performance on the Corr dataset. Across Fog, Frost, Motion, and Snow partitions, the BNN consistently provides lower risk at a given coverage level and higher accuracy (F1) for a given confidence threshold.}
    \label{fig:uq_corr}

\end{figure*}

We evaluated UQ performance using risk-coverage curves~\cite{ding2020revisiting}, where ``risk'' is defined as the probability of missing a victim ($1 -$\,Recall) and ``coverage'' is the percentage of detections retained after filtering by confidence.
In an autonomous SAR mission, the ability to filter detections by confidence is paramount; by discarding low-confidence predictions, the drone minimizes unnecessary verification maneuvers, thereby expanding the searchable area and decreasing rescue delay, which is a factor directly linked to the victim's chance of survival.
\hl{As shown in Fig.~\ref{fig:uq_sard}, which plots risk vs. coverage for the SARD dataset,} the BNN exhibits lower risk at every coverage level, reducing the area under the risk-coverage curve (AURC) by \qty{26.4}{\percent} ($0.3054$ to $0.2247$).
This indicates that the BNN can safely filter out more false positives without missing actual victims.
Calibration error, or the model's ability to approximate the empirical probability of a prediction being correct, was measured using adaptive binning to account for non-uniform confidence distributions~\cite{ding2020revisiting}.
The BNN improved expected calibration error (AECE) by \qty{56.2}{\percent} and maximum calibration error (AMCE) by \qty{36.9}{\percent}.
The reduction in AMCE is particularly critical for safety-critical SAR missions, as it 
ensures that rare but dangerous high-confidence errors are not masked by the average performance that AECE reports.
The imperfect CLT-GRNG output distribution degrades AURC by only \qty{0.49}{\percent} and increases AMCE by \qty{9.2}{\percent} compared to an ideal Gaussian distribution, a favorable trade-off for the massive energy gains.

\subsubsection{Robustness to Environmental Effects}

We further validated the models on the SARD ``Corr'' dataset, which synthetically introduces weather effects (fog, snow, frost) and motion blur to simulate adverse flight conditions.
\hl{As shown by the risk-coverage plots and F1-confidence plots of Fig.~\ref{fig:uq_corr}, the BNN exhibits less risk across the coverage spectrum and higher F1 across the model confidence spectrum for every corruption.
Table~\ref{tab:aurc} summarizes the performance across all discussed datasets.}
On average, the CNN lost \qty{19.2}{\percent} accuracy (mAP-50) in the corrupted datasets, but the BNN provides more robustness to environmental effects at an average loss of \qty{14.5}{\percent}.
Like in the SARD dataset, the imperfect CLT-GRNG output distribution does not significantly impact accuracy or uncertainty quantification performance for the corrupted datasets; average mAP-50 negligibly decreases (\qty{-0.02}{\percent}), and AURC, AECE, and AMCE all slightly improve (\qty{0.01}{\percent}, \qty{0.21}{\percent}, and \qty{0.89}{\percent}, respectively), though this effect is within the margin of error for this study.
Across all corruption types, the CLT-GRNG BNN improved average mAP-50 by \qty{6.0}{\percent}, AURC by \qty{14.4}{\percent}, AECE by \qty{32.8}{\percent}, and AMCE by \qty{22.1}{\percent} compared to the CNN.
This confirms that the proposed hardware maintains the BNN's intrinsic ability to flag out-of-distribution data, preventing confident failures in hazardous weather.

\begin{table}[]
    \centering
    \caption{Summary of Model Accuracy and UQ Performance}
    \label{tab:aurc}
    \begin{tabular*}{\columnwidth}{@{\extracolsep{\fill}} cc|cccc }
    \toprule
        \textbf{Partition}\
        & \textbf{Model}\
        & \textbf{mAP-50}$\uparrow$\
        & \textbf{AURC}$\downarrow$\
        & \textbf{AECE}$\downarrow$\
        & \textbf{AMCE}$\downarrow$\\

        \midrule

        Fog\
        & CNN\
        & \textcolor{red4}{\textbf{0.6972}}\
        & \textcolor{red4}{\textbf{0.4741}}\
        & \textcolor{red4}{\textbf{0.2146}}\
        & \textcolor{red4}{\textbf{0.4695}}\\

        Fog\
        & BNN\
        & \textcolor{green5}{\textbf{0.7542}}\
        & \textcolor{green5}{\textbf{0.4113}}\
        & 0.1702\
        & \textcolor{green5}{\textbf{0.4126}}\\

        Fog\
        & This$^\mathrm{\ast}$\
        & 0.7531\
        & 0.4120\
        & \textcolor{green5}{\textbf{0.1697}}\
        & \textcolor{green5}{\textbf{0.4126}}\\

        \midrule

        Frost\
        & CNN\
        & \textcolor{red4}{\textbf{0.5845}}\
        & \textcolor{red4}{\textbf{0.5470}}\
        & \textcolor{red4}{\textbf{0.1934}}\
        & \textcolor{red4}{\textbf{0.3979}}\\

        Frost\
        & BNN\
        & \textcolor{green5}{\textbf{0.6280}}\
        & \textcolor{green5}{\textbf{0.4644}}\
        & \textcolor{green5}{\textbf{0.1195}}\
        & \textcolor{green5}{\textbf{0.2849}}\\

        Frost\
        & This$^\mathrm{\ast}$\
        & 0.6275\
        & 0.4646\
        & \textcolor{green5}{\textbf{0.1195}}\
        & \textcolor{green5}{\textbf{0.2849}}\\

        \midrule

        Motion\
        & CNN\
        & \textcolor{red4}{\textbf{0.7963}}\
        & \textcolor{red4}{\textbf{0.4188}}\
        & \textcolor{red4}{\textbf{0.1477}}\
        & \textcolor{red4}{\textbf{0.3160}}\\

        Motion\
        & BNN\
        & 0.8280\
        & 0.3770\
        & 0.1057\
        & \textcolor{green5}{\textbf{0.2688}}\\

        Motion\
        & This$^\mathrm{\ast}$\
        & \textcolor{green5}{\textbf{0.8291}}\
        & \textcolor{green5}{\textbf{0.3762}}\
        & \textcolor{green5}{\textbf{0.1052}}\
        & \textcolor{green5}{\textbf{0.2688}}\\

        \midrule

        Snow\
        & CNN\
        & \textcolor{red4}{\textbf{0.6920}}\
        & \textcolor{red4}{\textbf{0.4718}}\
        & \textcolor{red4}{\textbf{0.1634}}\
        & \textcolor{red4}{\textbf{0.3515}}\\

        Snow\
        & BNN\
        & 0.7258\
        & 0.3837\
        & \textcolor{green5}{\textbf{0.0889}}\
        & 0.2394\\

        Snow\
        & This$^\mathrm{\ast}$\
        & \textcolor{green5}{\textbf{0.7258}}\
        & \textcolor{green5}{\textbf{0.3834}}\
        & \textcolor{green5}{\textbf{0.0889}}\
        & \textcolor{green5}{\textbf{0.2287}}\\

        \midrule

        SARD\
        & CNN\
        & \textcolor{red4}{\textbf{0.8573}}\
        & \textcolor{red4}{\textbf{0.3054}}\
        & \textcolor{red4}{\textbf{0.0553}}\
        & \textcolor{red4}{\textbf{0.1826}}\\

        SARD\
        & BNN\
        & 0.8588\
        & 0.2247\
        & \textcolor{green5}{\textbf{0.0242}}\
        & \textcolor{green5}{\textbf{0.1153}}\\

        SARD\
        & This$^\mathrm{\ast}$\
        & \textcolor{green5}{\textbf{0.8607}}\
        & \textcolor{green5}{\textbf{0.2236}}\
        & 0.0246\
        & 0.1259\\

        \bottomrule

    \end{tabular*}
    {\raggedright \newline $^\mathrm{\ast}$BNN using the CLT-GRNG output distribution \par}
\end{table}

%% file: 05_Conclusion.tex
\section{Conclusion}
\label{sec:conclusion}

This work addresses a critical bottleneck in autonomous aerial SAR: the tendency of deterministic deep learning models to generate high-confidence false positives.
These errors compel drones to expend limited energy and flight time investigating invalid targets---delays that directly reduce the searchable area and the probability of victim survival.
While Bayesian neural networks provide the uncertainty quantification necessary to filter these anomalies and adapt to dynamic environments, their computational demands have historically precluded deployment on battery-constrained edge devices.
We bridge this gap with a novel FeFET-based accelerator driven by a write-free, central limit theorem-based GRNG (CLT-GRNG).
Unlike prior stochastic architectures that rely on energy-intensive write cycles or large-area devices to shape distributions, our approach dynamically aggregates currents from minimum-sized, pre-programmed FeFETs.
This strategy eliminates the latency and endurance penalties of reprogramming, ensuring the system can sustain long-duration missions.
The resulting accelerator achieves an \hl{area-normalized tile energy efficiency} of \qty{185}{\text{TOPS/W/}\mm\squared} with the CLT-GRNG consuming only \qty{640}{\atto\joule} per sample.
Crucially, this massive reduction in energy comes with a negligible \qty{1.7}{\percent} increase in expected calibration error, demonstrating that robust, uncertainty-aware DL is now viable within the strict power envelopes of next-generation rescue fleets.